\documentclass[12pt,a4paper]{article}
\usepackage{amssymb,enumitem,booktabs,subfig,xcolor,microtype,setspace,bm,longtable,paralist,dsfont,fancyvrb}
\usepackage[top=1in, bottom=1in, left=1in, right=1in]{geometry}
\usepackage{natbib,url,palatino,eulervm}
\usepackage[pdftex,colorlinks=true,hypertexnames=false]{hyperref}
\definecolor{darkblue}{rgb}{0,0,.6}
\hypersetup{citecolor=darkblue,linkcolor=darkblue,urlcolor=darkblue}
\usepackage{amsmath,orcidlink,rotating,tikz,amsfonts,epsfig}
\usepackage{graphics,graphicx,lscape}
\usepackage[normalem]{ulem}
\usetikzlibrary[decorations.shapes]
\usetikzlibrary[petri]
\usetikzlibrary{positioning}
\usepackage[linewidth=1pt]{mdframed}
\allowdisplaybreaks[4]
\DeclareMathOperator*{\argmin}{arg\,min}

\usepackage[font=small]{caption}

\providecommand{\U}[1]{\protect\rule{.1in}{.1in}}
\renewcommand{\baselinestretch}{1.2}
\graphicspath{{plots/}}

\usepackage{amsthm,thmtools}
\usepackage{mathrsfs}

\makeatletter
\def\th@newremark{\th@remark\thm@headfont{\bfseries}}
\makeatletter
\theoremstyle{newremark}

\newcommand{\X}{\mathcal{X}}
\newcommand{\Y}{\mathcal{Y}}
\newcommand{\F}{\mathcal{F}}
\declaretheoremstyle[
  spaceabove=6pt, spacebelow=6pt,
  headfont=\bfseries,
  notefont=\mdseries, notebraces={(}{)},
bodyfont=\normalfont,
  postheadspace=0.5em,
]{mystyle}

\makeatletter
\newcommand*{\addFileDependency}[1]{
\typeout{(#1)}
\@addtofilelist{#1}
\IfFileExists{#1}{}{\typeout{No file #1.}}
}\makeatother

\newcommand{\Rlogo}{\protect\includegraphics[height=1.8ex,keepaspectratio]{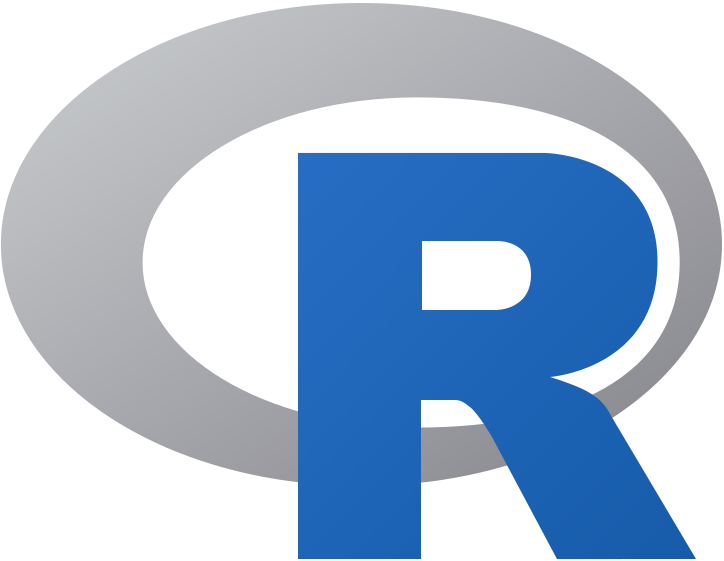}}

\begin{document}

\def\spacingset#1{\renewcommand{\baselinestretch}
{#1}\small\normalsize} \spacingset{1}

\title{\bf Forecasting intraday particle number size distribution: \mbox{A functional time series approach}}
\author{
\normalsize  Han Lin Shang \orcidlink{0000-0003-1769-6430} \\
\normalsize  Department of Actuarial Studies and Business Analytics \\
\normalsize  Macquarie University \\
\\
\normalsize Israel Martinez Hernandez \orcidlink{0000-0002-4122-2529}\\
\normalsize School of Mathematical Sciences \\
\normalsize Lancaster University
}
\date{}
\maketitle

\begin{abstract}
Particulate matter data now include various particle sizes, which often manifest as a collection of curves observed sequentially over time. When considering 51 distinct particle sizes, these curves form a high-dimensional functional time series observed over equally spaced and densely sampled grids. While high dimensionality poses statistical challenges due to the curse of dimensionality, it also offers a rich source of information that enables detailed analysis of temporal variation across short time intervals for all particle sizes. To model this complexity, we propose a multilevel functional time series framework incorporating a functional factor model to facilitate one-day-ahead forecasting. To quantify forecast uncertainty, we develop a calibration approach and a split conformal prediction approach to construct prediction intervals. Both approaches are designed to minimise the absolute difference between empirical and nominal coverage probabilities using a validation dataset. Furthermore, to improve forecast accuracy as new intraday data become available, we implement dynamic updating techniques for point and interval forecasts. The proposed methods are validated through an empirical application to hourly measurements of particulate matter in 51 size categories in London.

\vspace{.1in}
\end{abstract}

\newpage
\spacingset{1.48}

\section{Introduction}\label{sec:intro}

Modelling and forecasting particulate matter is crucial in environmetrics because of its implications for public health and policy making. As hundreds of millions of people worldwide experience the impact of climate change, it takes an incalculable toll on human health and well-being. Several statistical methods have been proposed to forecast particle matters \citep[see][for review]{DYK+09}. These include neural networks \citep{PKK+11} and multiple linear modelling \citep{SHP08}, where a comparison between these two methods is presented in \cite{SKK06}. A commonality in these works is that they often employ discrete-time models, which depend on the specific time points at which measurements are taken. Since the concentrations of the particle matter are analysed as a discrete time series, we cannot recover the underlying continuous stochastic process that generates these observations. Instead, we can consider a functional data-analytic approach to analyse the changes in shape over time. Unlike previous studies that focus on a single particle size, our work considers a range of particle sizes.

In Section~\ref{sec:2}, we describe the daily curves of the hourly concentration of particle number size distribution with various aerodynamic diameters, ranging from $16.55$ to $604.4$ nanometres (nm), cf. 2500nm = 2.5um, which is the size of PM$_{2.5}$. By treating data as a time series of functions, we resort to functional data analysis of \cite{RS05} to extract information and to represent the intraday dependence. Let $P_{t}^{s}(u_j)$, $t=1,\dots,n$, $j=1,\dots,p$, $s=1,\dots,S$ be the concentration of particle matter at the intraday time $u_j$ on day $t$ of size $s$. To stabilise the variance and better normalise the variance, we take the $\log_{10}$ transformation \citep[see also][]{GKR17, IMH23}, i.e., we work with the data
\begin{align}
\X_t^s(u_j) &= \log_{10}[P_t^s(u_j)+1] \\
P_t^s(u_j) & = 10^{\X_t^s(u_j)} - 1, \label{eq:inverse_transformation}
\end{align}
where one was added to avoid a zero count. By linear interpolation, we consider intraday curves as continuous curves $\X_t^s(u)$ constructed from $\X_t^s(u_j)$, $j=1,\dots,p$, where $p$ denotes the total number of intraday measurements of particle counts of size $s$. Since our data are equally spaced and densely observed, a linear interpolation of \cite{HAB+25} is computationally efficient for the data \citep[see also][]{ZW16, CTC+24}. In our example, there are $p=24$ hourly time intervals representing 24 hours of intraday measurements. Once we have constructed a time series of functions, we work directly with a set of functional time series, denoted by $[\X_1^s(u),\dots,\X_n^s(u)]$, for $u\in \mathcal{I}$ where $\mathcal{I}$ denotes a function support range.

In the statistical literature, there has been a surge of interest in the development of (univariate) functional time series forecasting methods \citep[see][for a general background]{HK12, KR17}. From a parametric perspective, \cite{Bosq00} propose the functional autoregressive of order one and derive one-step-ahead forecasts that are based on a regularised form of the Yule-Walker equations. \cite{KR13} extend functional autoregressive of order one to a higher order and propose a selection criterion to determine the optimal order. \cite{KK17} propose the functional moving average model and introduce an innovations algorithm to obtain the best linear predictor. \cite{KKW17} propose the functional autoregressive moving average process to predict highway traffic flows. From a nonparametric perspective, \cite{HS09} apply functional principal component analysis to decompose a time series of functions into a set of functional principal components and their associated scores. While \cite{HS09} use a univariate time-series forecasting method to forecast each set of principal component scores, \cite{ANH15} consider a multivariate time-series forecasting method. Although the multivariate time-series forecasting method can capture the cross-correlation among scores, the univariate time-series forecasting method can model nonstationarity within each set of scores. Using historical curves and estimated functional principal components, the point forecasts are obtained by multiplying the forecast scores with estimated functional principal components.

The presence of multiple functional time series is not uncommon, such as subnational mortality rates in \cite{JSS24, JSS25}, firm-specific cumulative intraday return in \cite{LLS+25}, half-hourly energy consumption readings for thousands of London households \citep{CFQ+25}, or hourly wind speed (m/s) over 5-km resolution in Saudi Arabia \citep{MG23}. In Section~\ref{sec:3}, we implement a multilevel functional time series method of \cite{Shang16} to produce \textit{one-day-ahead} forecasts for a range of particle sizes. As a joint modelling technique, the multilevel functional time series method resembles a two-way functional analysis of variance studied by \cite{MVB+03}, and it is a special case of the general ``functional mixed model'' in \cite{MC06}.

When a functional time series is constructed from segments of a longer univariate time series, the most recent curve is observed sequentially and thus may not represent a complete curve. When we observe the first $m_0$ time periods of $\X_{n+1}^s(u)$, denoted by $[\X_{n+1}^s(u_e)=[\X_{n+1}^s(u_1),\dots,\X_{n+1}^s(u_{m_0})]^{\top}$, we are interested in forecasting the data in the remaining period of day $n+1$, denoted by $\X_{n+1}^s(\mathsf{u})$, where $\mathsf{u}\in \mathcal{I}_l$ denotes the function support range for the updating period and $s=1, \dots, S$ represents different sizes of particles. The one-day-ahead forecasts do not utilise the newly arrived intraday observations, so it is desirable to dynamically update the point forecasts for the remaining time period of the most recent day $n+1$. In Section~\ref{sec:4}, we consider several dynamic updating methods to improve point and interval forecast accuracies for different sizes of particles \citep[see also][for function-on-function regression approaches]{Chiou12, JAO13}. In function-on-function regression, the predictor is a block corresponding to newly observed data, while the response is another block corresponding to the data in the update period.

In Section~\ref{sec:5}, we present two ways to construct pointwise prediction intervals. Both approaches are distribution-free, model-agnostic, and computationally fast, since they only rely on a set of validation data. When newly arrived data are observed, they can also be used to update interval forecasts. Using the point and interval forecast error metrics in Section~\ref{sec:6}, we evaluate and compare the point and interval forecast accuracy in Section~\ref{sec:7}. The conclusion is presented in Section~\ref{sec:8}, along with some ideas on how the methodology can be further extended.

\section{Intraday particle number size distribution (PNSD)}\label{sec:2}

We consider particle number size distribution data, which provide information on particulate matter sizes and are widely used due to the richness of information they contain. PNSD involves considering particle sizes that range from $10$ to $2,500$ nm, encompassing the fine and ultrafine domains. Furthermore, observations can be recorded at a fine time resolution, resulting in high-dimensional time series, where the dimension is the number of different sizes in the PNSD data. The PNSD data to be analysed in this article were collected at a single urban background monitoring station located in a residential area of London (North Kensington, 51.52 N, 0.21 W; 27m above sea level). The station is part of the London Air Quality Network and the National Automatic Urban and Rural Network. It is situated on the grounds of Sion Manning School in St Charles Square. The raw data consist of hourly measurements from January 2011 to August 2018, obtained with an SMPS TSI 3080 + CPC TSI 3775, fitted with a long DMA instrument and a diffusion dryer, following the recommendations of the EUSAAR/ACTRIS protocol \citep{WBNS2012}. At each time point $u_j$ (hr), $51$ different particle sizes are measured \footnote{Strictly speaking, $51$ is the number of size categories considered.}, ranging from $16.55$ to $604.4$ nm; $P_t^{1}(u_j), \ldots, P_t^{51}(u_j)$. Thus, for each day, the raw data is a $24\times 51$ matrix. Data can be downloaded from the \href{https://uk-air.defra.gov.uk/networks/site-info?site_id=KC1}{Department for Environment, Food and Rural Affairs} webpage. Also, see \cite{IMH2024RSSc}, where the same data were analysed but to estimate the source profiles rather than forecasting.  

\begin{figure}[!htb]
\centering
{\includegraphics[width=8.72cm]{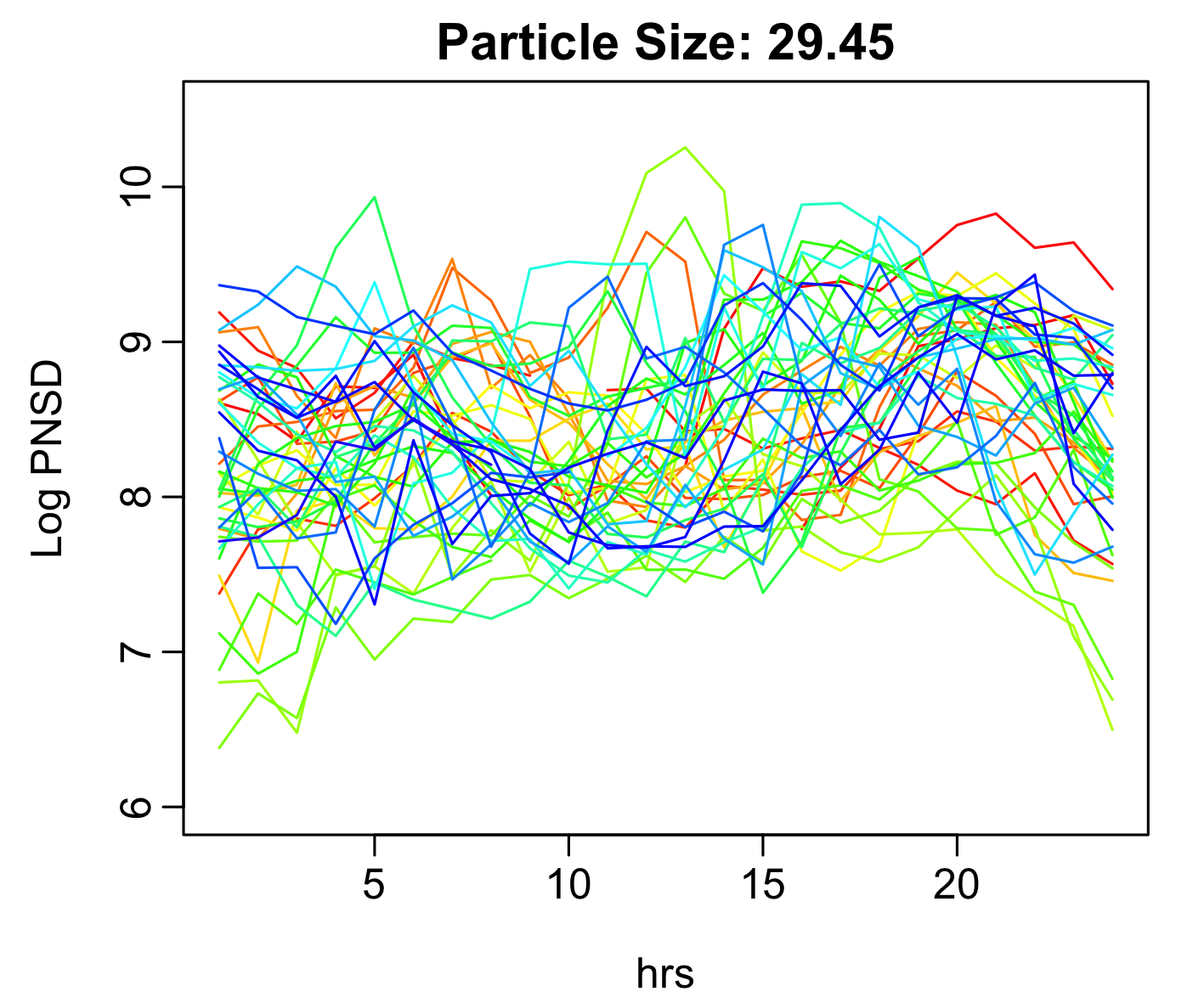}}
\quad
{\includegraphics[width=8.72cm]{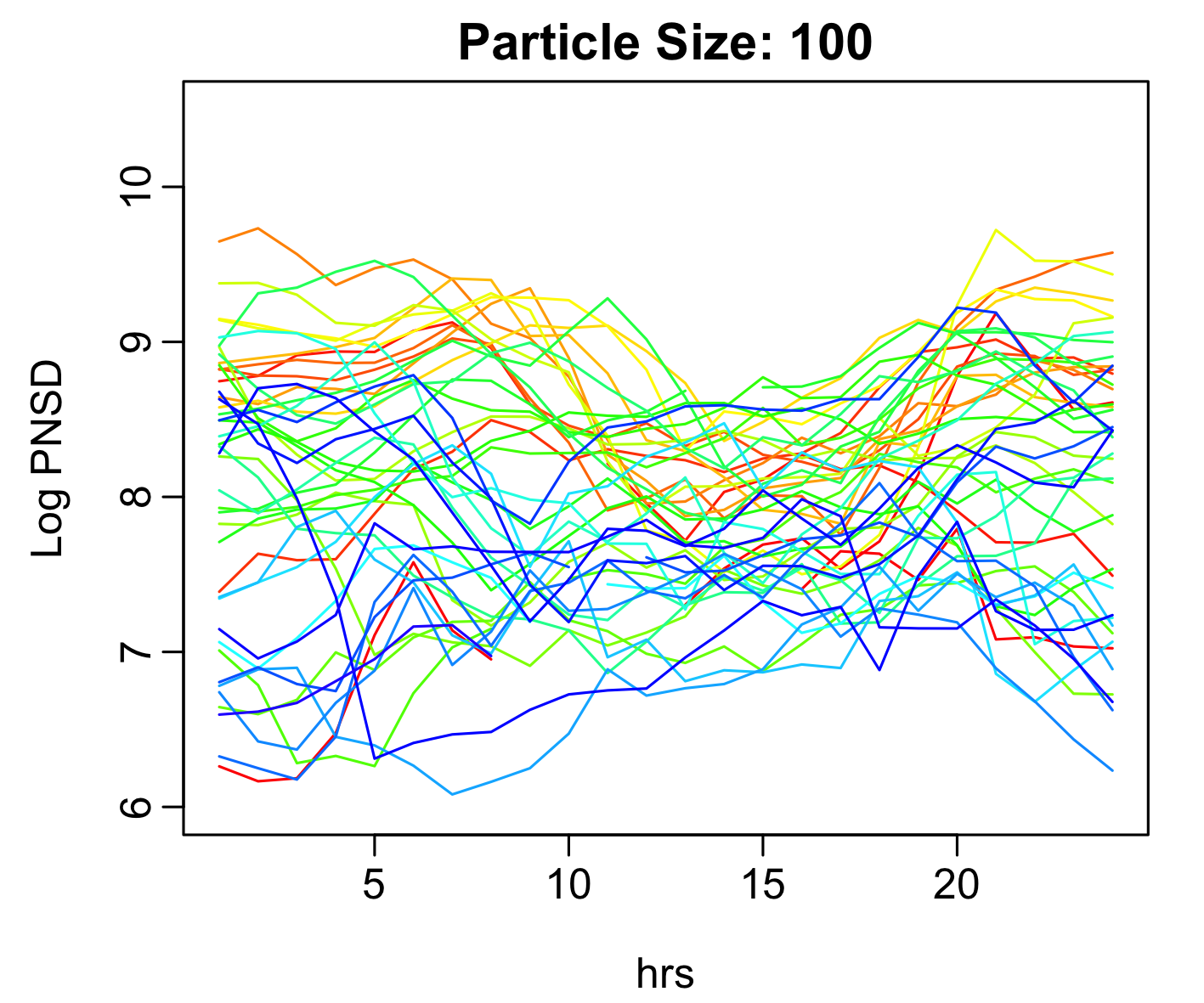}}
\caption{Functional time series of the PNSD for two sizes. Each plot illustrates $50$ consecutive days of data, where each function represents one day. We can observe that the evolution of the time series of functions depends on the size.}\label{fig:s2-1}
\end{figure}

We treat PNSD data as multivariate time series of functions ($51$-dimensional multivariate functional data). For each day~$t$, we assume that we have $51$ continuous functions instead of a matrix. Then, we apply the $\log_{10}$ transformation, as described in the introduction, to obtain $\X_t^s (u_j)= \log_{10}[P_t^s(u_j)+1]$, where one was added to avoid zero count. Finally, the continuous function is obtained using interpolation: $\X_t^1(u), \X_t^2(u),\ldots, \X_t^{51}(u)$, where $\X_t^s(u)$ represents the continuous intraday log PNSD measurement for size $s$. Figure~\ref{fig:s2-1} shows two functional time series corresponding to particle sizes $29.45$ nm and $100$ nm. Each plot presents only $50$ consecutive days. Initially, we can observe that each functional time series exhibits a distinct dynamic and trend. In our model, we will consider the different variabilities between different sizes and obtain the one-day-ahead point and interval forecasts.

The data exhibit an inherent weekly periodicity. For example, consecutive Sundays tend to be more similar to each other than two consecutive days, such as Saturday and Sunday. This similarity arises from people's routines; for example, there is typically less traffic on weekends compared to weekdays. As a result, fewer particles are emitted from vehicle emissions during this time. Therefore, we define our time series of continuous functions as a sequence based on the days of the week, such as the sequence of all Mondays. With this approach, the sample size of the multivariate functional time series is 408 (corresponding to the total number of weeks in the data). Although it is worth noting that this approach can easily be modified to accommodate weekly curves or even monthly curves. In the forecasting experiment, we also consider weekly curves from 168 data points. Using the functional time-series method, we produce one-week-ahead forecasts, and these forecasts are then segmented according to each day of the week.
\begin{figure}[!htb]
\centering
\includegraphics[width=12cm]{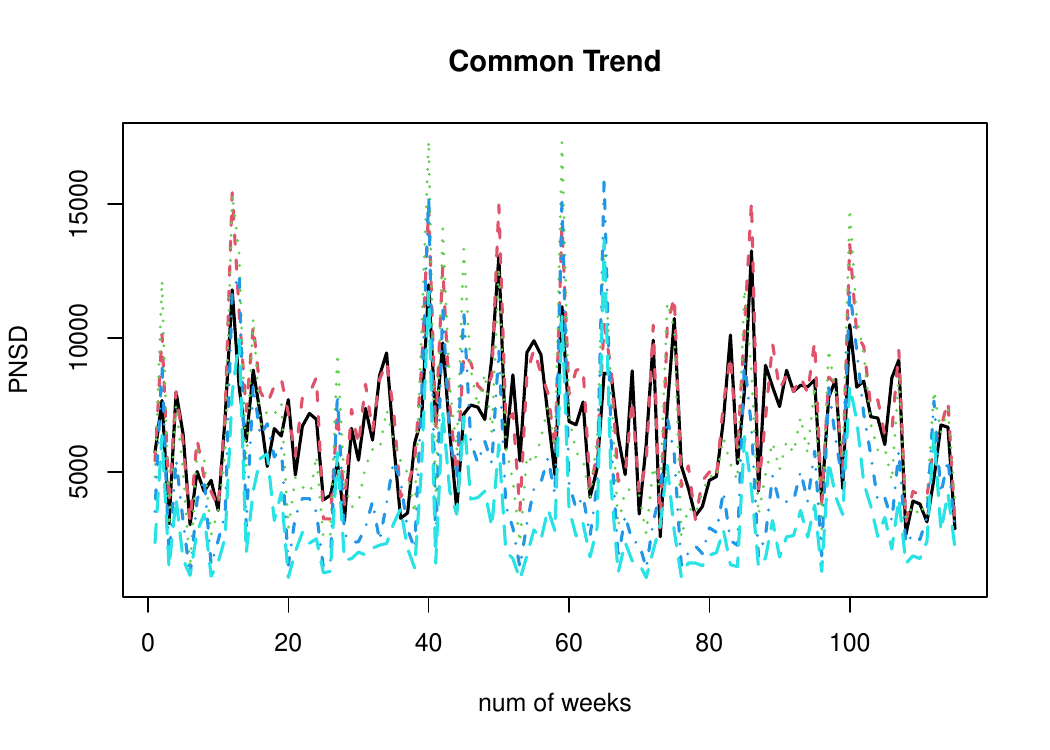}
\caption{Time series of average daily counts for a given day (sequence of Mondays). Each time series represents different sizes (in nm): $29.45,  39.25,  64.95, 100.00, 133.40$. We can see that there is a common trend.}\label{fig:trend}
\end{figure}

Given the strong correlation among different particle sizes, it is expected that a common trend will be observed - denoted as $R_t(u)$ in Section~\ref{sec:3}. One way to visualise this common trend is as follows. Given the sequence of functional data, for each size $s$, we can average the counts for each functional data, $(1/24) \sum_{j=1}^{24} P^s_t(u_j)$, for all $t$, resulting in a univariate time series. This time series allows us to visualise the trend over time. Figure~\ref{fig:trend} shows five time series corresponding to particle sizes (nm) $s=29.45,  39.25,  64.95, 100.00, 133.40.$ We can see that there is evidence of a common trend since all time series have a similar pattern. Therefore, we will consider a common trend component in our model, defined in Section~\ref{sec:3}.

\begin{figure}[!htb]
\centering
\includegraphics[width=12cm]{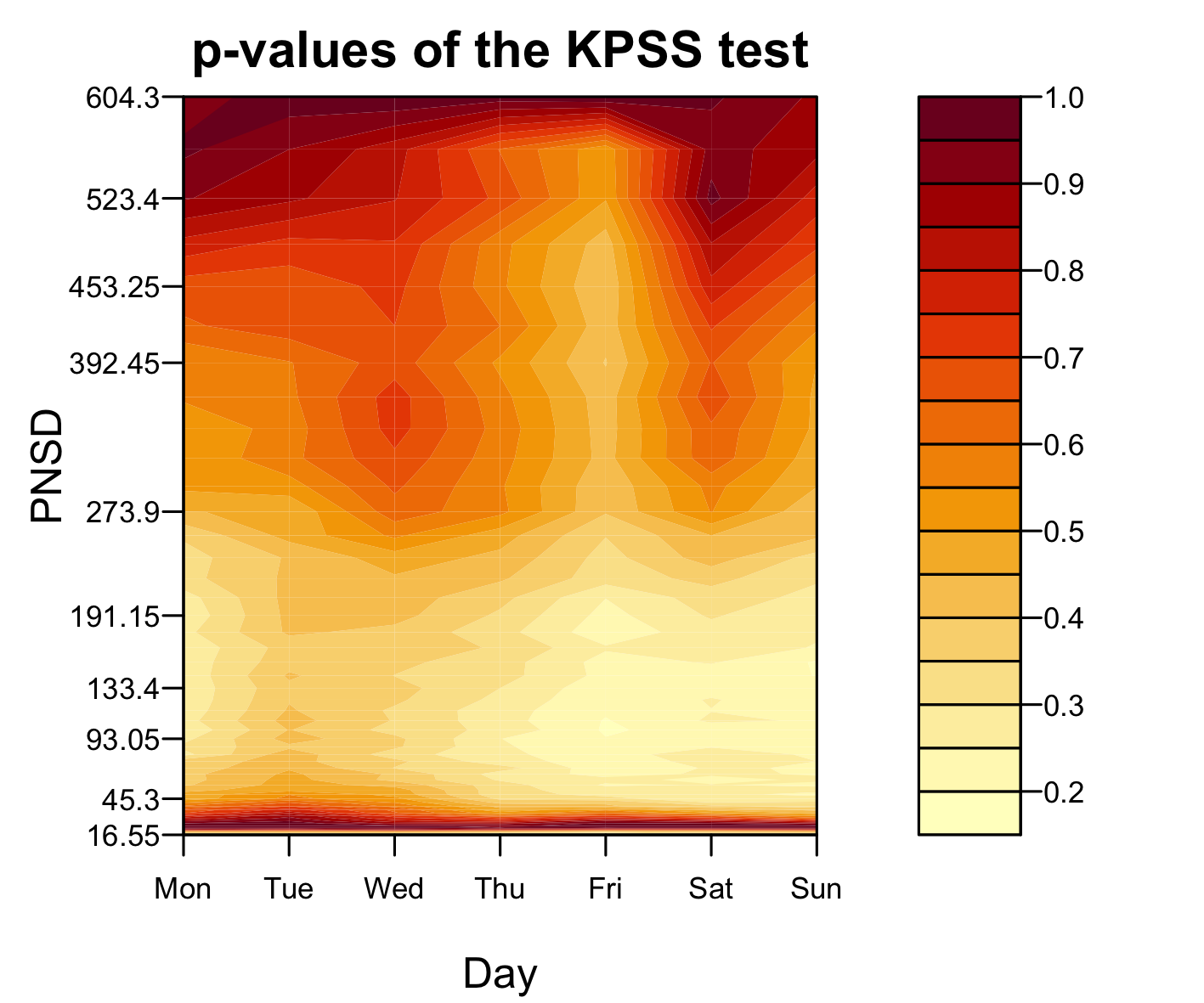}
\caption{An image plot containing the $p$-values of the functional KPSS test of \cite{HKR14} for each series.}\label{fig:KPSS}
\end{figure}

To examine stationarity, we implement a functional KPSS test of \cite{HKR14} for each of the 51 series. From the $p$-value image plot in Figure~\ref{fig:KPSS}, we conclude that all series are stationary since the $p$-value $>0.1$.

\section{Multiple functional time series forecasting method}\label{sec:3}

\subsection{Multilevel functional time-series (MLFTS) method}\label{sec:3.1}

The essence is to decompose curves among different sizes into a size-specific mean $\mu^s(u)$, a common trend among sizes $R_t(u)$, a size-specific residual trend $U_t^s(u)$, and measurement error $e_t^s(u)$. The common trend and size-specific residual trends are modelled by projecting them onto the eigenvectors of the covariance operators of the aggregate and size-specific centred stochastic processes, respectively. The aggregated series can be formed by averaging functional time series in various sizes. To be specific, $\X_t^s(u)$ can be written as
\begin{equation*}
\X_t^s(u) = \mu^s(u)+R_t(u)+U_t^s(u)+e_t^s(u),\qquad u\in \mathcal{I}.
\end{equation*}
The size-specific mean is estimated by $\widehat{\mu}^s(u) = \frac{1}{n}\sum^{n}_{t=1}\X_{t}^{s}(u)$, the common and size-specific trends can be estimated by
\begin{align*}
\widehat{R}_t(u) = \sum^{\infty}_{k=1}\widehat{\beta}_{t,k}\widehat{\phi}_k(u) \approx \sum_{k=1}^K\widehat{\beta}_{t,k}\widehat{\phi}_k(u) \\
\widehat{U}_t^s(u) = \sum^{\infty}_{l=1}\widehat{\gamma}^s_{t,l}\widehat{\psi}_l^s(u) \approx \sum_{l=1}^L\widehat{\gamma}_{t,l}^s\widehat{\psi}_l^{s}(u),
\end{align*}
where $\widehat{\bm{\beta}}_k=(\widehat{\beta}_{1,k},\dots,\widehat{\beta}_{n,k})$ represents the $k$\textsuperscript{th} principal component scores of $\bm{R}(u)=[R_1(u),\dots,R_n(u)]$, and $\widehat{\bm{\Phi}}(u) = [\widehat{\phi}_1(u),\dots,\widehat{\phi}_K(u)]$ are the corresponding functional principal components. Similarly, $\widehat{\bm{\gamma}}^s_{l}=(\widehat{\gamma}^s_{1,l},\dots,\widehat{\gamma}^s_{n,l})$ represents the $l$\textsuperscript{th} principal component scores of $\bm{U}^s(u)=[U_1^s(u),\dots,U_n^s(u)]$, and $\widehat{\bm{\Psi}}^s(u) = [\widehat{\psi}_1^s(u),\dots,\widehat{\psi}_n^s(u)]$ are the corresponding functional principal components. Averaged over 51 series, we compute the common trend as $R_t(u) = \frac{1}{51}\sum^{51}_{s=1}\X_t^s(u)$. The series-specific trend is captured by $U_t^s(u) = \X_t^s(u) - R_t(u)$. To model $R_t(u)$ and $U_t^s(u)$, we resort to functional principal component analysis of their covariance functions, respectively. So, $\X_t^s(u)$ can be estimated by
\begin{equation}
\X_t^s(u) = \widehat{\mu}^s(u) + \sum_{k=1}^K\widehat{\beta}_{t,k}\widehat{\phi}_k(u) + \sum_{l=1}^L\widehat{\gamma}_{t,l}^s\widehat{\psi}_l^{s}(u) + e_t^s(u). \label{eq:MLFTS}
\end{equation}
Notice that the intraday dependence is described with the covariance operators obtained with $\bm{R}(u)$ and $\bm{U}^s(u)$.

From the estimated eigenvalues, we can estimate the proportion of variability explained by the aggregated data; a measure of within-cluster variability is given by
\begin{equation}
\frac{\int_{u\in\mathcal{I}} \text{Var}[\bm{R}(u)]du}{\int_{u\in\mathcal{I}} \text{Var}[\bm{R}(u)]du + \int_{u\in \mathcal{I}} \text{Var}[\bm{U}^s(u)]du}\approx \frac{\sum_{l=1}^L\lambda_l}{\sum_{l=1}^L\lambda_l+\sum^M_{m=1}\lambda_m}, \label{eq:0}
\end{equation}
where $\lambda_l$ and $\lambda_m$ denote the population eigenvalues of the common trend and the size-specific residual trend. Equation~\eqref{eq:0} not only presents a modelling and forecasting approach but also allows us to better understand and interpret the shared variability.

The selection of retained functional principal components has garnered considerable attention. Some popular estimation methods include:
\begin{inparaenum}[(1)]
\item scree plots or the fraction of variance explained by the first few functional principal components \citep{Chiou12};
\item Akaike information criterion \citep{ANH15} and Bayesian information criterion \citep{OS24};
\item predictive cross-validation with one-curve-leave-out \citep{RS91};
\item the bootstrap technique \citep{HV06};
\item eigenvalue ratio criterion \citep{LRS20};
\item set $K=L=6$ \citep{HBY13}. 
\end{inparaenum}
Here, we consider the simplest strategy by setting $K=L=6$. From a forecast accuracy viewpoint, selecting more than the optimal number of components does not harm the forecast accuracy. Since the eigenvalues are ordered in a decreasing order, the forecasts of the additional principal component scores tend to close to zero.

By conditioning on the estimated functional principal components $[\widehat{\phi}_1(u),\dots,\widehat{\phi}_K(u)]$ and $[\widehat{\psi}_1^s(u),\dots,\widehat{\psi}_L^s(u)]$ in~\eqref{eq:MLFTS} and observed data, the one-step-ahead point forecast can be obtained as
\begin{equation*}
\widehat{\X}_{n+1|n}^s(u) = \widehat{\mu}^s(u) + \sum_{k=1}^K\widehat{\beta}_{n+1|n,k}\widehat{\phi}_k(u) + \sum_{l=1}^L\widehat{\gamma}_{n+1|n,l}^s\widehat{\psi}_l^{s}(u),
\end{equation*}
where $\widehat{\beta}_{n+1|n,k}$ and $\widehat{\gamma}_{n+1|n,l}$ denote the one-step-ahead forecasts of the $k$\textsuperscript{th} and $l$\textsuperscript{th} principal component scores, respectively. These forecasts can be obtained from the exponential smoothing method \citep[see, e.g.,][]{HKO+08}. Using the inverse transformation in~\eqref{eq:inverse_transformation}, we obtain $\widehat{P}_{n+1|n}^s(u)$.

\subsection{Functional factor model}\label{sec:3.2}

To accommodate strong cross-sectional dependence driven by the latent structure, we may adopt the factor model approach to multiple functional time series. \cite{GSY19} apply the functional principal component analysis to each series and then model the component scores via the classic factor model. This method may lead to information loss due to two-stage dimension reduction. To avoid this issue, two types of functional factor model have been proposed with different constructions of common components. \cite{TNH23} introduce a high-dimensional functional factor model with functional loadings and scalar factors, while \cite{GQW24} propose a different version with scalar loadings and functional factors. \cite{LLS+25} propose a general framework expressed as
\begin{equation}
\X_t^s(u) = \sum_{\ell=1}^{L_{*}}\int_{v\in \mathcal{I}} B_{s\ell}(u,v)F_{t\ell}(v)dv+\varepsilon_{t}^s(u). \label{eq:1}
\end{equation}
For each component of the $\ell$-dimensional vector of functional factors, we consider the series approximation:
\begin{equation}
F_{t\ell}(v) = \Phi_{\ell}(v)^{\top}G_t + \eta_{t\ell}(v). \label{eq:2}
\end{equation}
By plugging~\eqref{eq:2} into~\eqref{eq:1}, we obtain
\begin{equation}
\X_t^s(u) = \sum_{\ell=1}^{L_{*}}\int_{v\in \mathcal{I}} B_{s\ell}(u,v)[\Phi_{\ell}(v)^{\top}G_t + \eta_{t\ell}(v)]dv+\varepsilon_{t}^s(u) \label{eq:3}
\end{equation}
Letting
\begin{equation}
\Lambda_s(u) = \sum_{\ell=1}^{L_{*}}\int_{v\in \mathcal{I}} B_{s\ell}(u,v)\Phi_{\ell}(v)^{\top}dv \label{eq:4}
\end{equation}
and
\begin{equation}
\Y_{t}^s(u) = \sum_{\ell=1}^{L_{*}}\int_{v\in \mathcal{I}} B_{s\ell}(u,v)\eta_{t\ell}(v)dv,\label{eq:5}
\end{equation}
combining~\eqref{eq:3}--\eqref{eq:5}, it leads to 
\begin{equation*}
\X_t^s(u) = \Lambda_s(u)^{\top}G_t+\underbrace{\Y_{t}^s(u) + \varepsilon_{t}^s(u)}_{\epsilon_{t}^s(u)}.
\end{equation*}

To estimate factor loading functions and real-valued factors, we resort to the least-squares objective function:
\begin{equation*}
\sum^n_{t=1}\sum^S_{s=1}\left\|\X_t^s(u) - \Lambda_s(u)^{\top}G_t\right\|_2^2,
\end{equation*}
where $\|\cdot\|_2$ denotes the norm of elements in the Hilbert space, $\Lambda_s(u)$ is a $q$-dimensional vector of functions and $G_t$ is a $q$-dimensional vector of numbers. Minimisation of the least-squares objective function can be solved via eigenanalysis of 
\begin{equation}
\bm{\Delta} = (\Delta_{tt^{\dagger}})_{n\times n} \quad \text{with} \quad \Delta_{tt^{\dagger}} = \frac{1}{S}\sum^S_{s=1}\int_{u\in \mathcal{I}} \X_{t}^s(u)\X_{t^{\dagger}}^s(u)du. \label{eq:6}
\end{equation}
Let $\widetilde{\bm{G}}=(\widetilde{G}_1, \widetilde{G}_2, \dots,\widetilde{G}_n)^{\top}$ be a matrix consisting of the eigenvector (multiplied by root-$n$) corresponding to the $q$ largest eigenvalues of $\bm{\Delta}$ in~\eqref{eq:6}. The factor loadings are estimated as
\begin{equation*}
\widetilde{\Lambda}_s(u) = \frac{1}{n}\sum^n_{t=1}\X_t^s(u)\widetilde{G}_t,
\end{equation*}
via least squares, using the normalisation restriction $\frac{1}{n}\sum^n_{t=1}\widetilde{G}_t\widetilde{G}_t^{\top} = \bm{I}_q$. To select $q$, we consider an information criterion \citep[see][equation 5.1]{LLS+25}. Consequently, the model residual function, itself a high-dimensional functional time series, can be expressed as $\widetilde{\epsilon}_t^s(u) = \X_t^s(u) - \widetilde{\Lambda}_s^{\top}(u)\widetilde{G}_t$. It can be modelled using the multilevel functional time-series method in Section~\ref{sec:3.1}.

\section{Updating point forecasts}\label{sec:4}

Although we present two functional time-series methods to produce one-day-ahead forecasts in Section~\ref{sec:3}, it is a feature of our problem where intraday measurements are continuously observed. Incorporating newly arrived information may improve forecast accuracy. In this section, we describe methods for incorporating partially observed curves into the model.

\subsection{Block moving (BM) method}\label{sec:4.1}

The BM method utilises functional time-series methods but redefines the beginning and end time points of curves. Since intraday time is a continuous variable, we can alter its function support from $[u_1, u_p]$ to $[u_1, u_{m_0}]\cup (u_{m_0}, u_p]$. With loss of some data points in the first curve highlighted in blue, a partially observed curve can be completed. In Figure~\ref{fig:1}, we present a concept diagram. 

\tikzset{decorate with/.style={fill=cyan!20,draw=cyan}}
\begin{figure}[!htb]
\begin{center}
\scalebox{.7}{
\begin{tikzpicture}
\draw (0,0) node[anchor=north east]{$u_p$} rectangle (12,8);
\draw (1.1,4.1) rectangle (13.1,12);
\draw (0, 8) node[anchor=north east]{$u_1$} -- (0, 0);
\draw (0,4.1) node[anchor=north east]{  \begin{rotate}{90} \hspace {-.5in} Dimensionality \end{rotate} \hspace{.05in} $u_{m_0}$} -- (1.1, 4.1);
\draw[dashed] (1.1,0) -- (1.1,4.1);
\draw[dashed] (2.2,12) -- (2.2,8);
\draw[dashed] (14.2, 12) -- (14.2, 8);
\draw (11.86, 8) node[anchor=south west]{$n+1$} -- (13, 8);
\draw[dashed] (13, 8) -- (14.2, 8);
\draw[dashed] (13, 12) -- (14.2, 12);
\draw[dashed] (13.1, 4.1) -- (13.1, 0);
\draw[dashed] (12, 0) -- (13.1, 0);
\draw (8.5,0) node[anchor=north east]{Number of curves} -- (12,0);
\draw(0.4,8) node[anchor=south west]{1} -- (1.1,8);
\draw [decorate with=rectangle] (0.0305,4.1365) -- (0.0305,7.98) --  (1.074,7.98) -- (1.074,4.1365);
\end{tikzpicture}}
\end{center}
\caption{Dynamic update via the block-moving approach. The coloured region shows the data loss on the first day. The forecasts for the rest of the day ($n+1$) can be updated using forecasts from a functional time-series method applied to the top block.}\label{fig:1}
\end{figure}
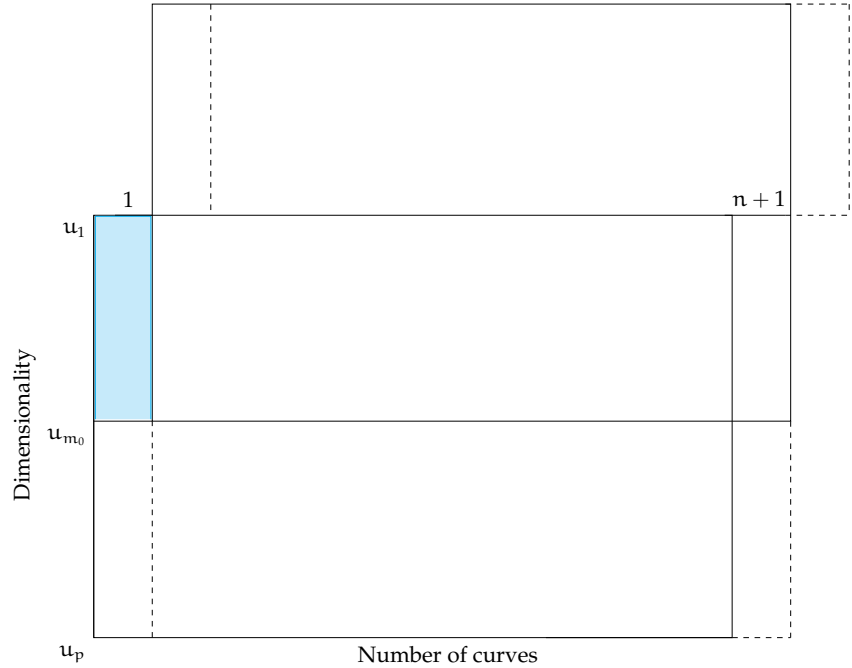

Note that the BM method is implemented $23$ times to cover different updating periods, from observing one hour of day $n+1$ to observing 23 hours of that day.

\subsection{Regression-based method}\label{sec:4.2}

\subsubsection{Ordinary least squares (OLS)}

The remaining part of the most recent curve can be updated using a regression-based approach based on a multivariate functional principal component analysis \citep[see, e.g.,][]{CCY14, SK22}. By stacking $s$, we can convert a three-dimensional array to a two-dimensional matrix, in which an eigenanalysis can be performed. Let $\bm{X}$ be a $(S\times p) \times n$ matrix, where $(S\times p)$ represents the stacked variables. From its empirical covariance of dimension $(S\times p)\times (S\times p)$, we obtain
\begin{equation*}
\bm{X} = \bm{F}\bm{B},
\end{equation*}
where $\bm{F}$ is a $(S\times p)\times N$ matrix, consisting of $N$ retained functional principal components, and $\bm{B}$ represents principal component scores of dimension $N\times n$. The advantage of multivariate functional principal component analysis is its ability to model the correlation among~$S$ particulate matter sizes.

For a given particle size $s$, let $\bm{\mathcal{F}}$ be a $p\times N$ matrix and $\bm{\F}_e$ be the $m_0\times N$ matrix of the explanatory variable, whose $(i,j)$\textsuperscript{th} entry is $\widehat{\phi}_i^s(u_j)$ for $1\leq j\leq m_0$ and $1\leq i\leq N$. Let $\X_{n+1}^s(u_e)$ be the $m_0\times 1$ matrix of the response variable, $\bm{\beta}_{n+1}^s=(\beta^s_{n+1,1},\dots,\beta^s_{n+1,N})^{\top}$ be the regression coefficients, and $\eta^s_{n+1}(u_e) = [\eta^s_{n+1}(u_1),\dots,\eta^s_{n+1}(u_{m_0})]^{\top}$ be the error term.  When the mean-adjusted $\widehat{\X}_{n+1}^{*,s}(u_e) = \X_{n+1}^s(u_e) - \widehat{\mu}^s(u_e)$ becomes available, the regression equation can be expressed as
\begin{equation*}
\widehat{\X}_{n+1}^{*,s}(u_e) = \bm{\F}^s_e\bm{\beta}^s_{n+1}+\eta^s_{n+1}(u_e).
\end{equation*}
The $\bm{\beta}^s_{n+1}$ can be estimated via an OLS estimator, giving
\begin{equation*}
\widehat{\bm{\beta}}_{n+1}^{s, \text{OLS}} = ((\bm{\F}_e^{s})^{\top}\bm{\F}^s_e)^{-1}(\bm{\F}_e^{s})^{\top}\widehat{\X}_{n+1}^{*,s}(u_e).
\end{equation*}
The OLS forecast of $\X_{n+1}^s(u_l)$ can be given by
\begin{equation*}
\widehat{\X}_{n+1}^{s, \text{OLS}}(u_l) = \widehat{\mu}^s(u_l) + \sum^N_{i=1}\widehat{\beta}_{n+1,i}^{s, \text{OLS}}\widehat{\phi}_i^s(u_l),
\end{equation*}
where $\{u^s_{m_0+1},\dots, u^s_{\tau}; \tau\leq p\}$ represents the discretised time points in the remaining intraday time period.

\subsubsection{Ridge regression (RR)}

The OLS estimator can be numerically unstable due to the use of the pseudo-inverse. To solve this problem, we employ the ridge regression estimator proposed by \cite{HK70} by introducing a penalty. The RR method shrinks the regression coefficient estimates towards zero. In statistics, it is also known as Tikhonov regularisation \citep[see, e.g.,][]{BCF17}. It aims to minimise a penalised residual sum of squares:
\begin{equation*}
\argmin_{\bm{\beta}^s_{n+1}}\left\{(\widehat{\X}_{n+1}^{*,s}(u_e) - \bm{\F}^s_e\bm{\beta}^s_{n+1})^{\top}(\widehat{\X}_{n+1}^{*,s}(u_e) - \bm{\F}^s_e\bm{\beta}^s_{n+1})+\lambda (\bm{\beta}_{n+1}^s)^{\top}\bm{\beta}^s_{n+1}\right\},
\end{equation*}
where $\lambda>0$ is a shrinkage parameter that controls the amount of shrinkage. Taking the first-order derivative with respect to $\bm{\beta}_{n+1}$, the RR estimate is 
\begin{equation*}
\widehat{\bm{\beta}}_{n+1}^{s, \text{RR}} = ((\bm{\F}_e^{s})^{\top}\bm{\F}_e+\lambda\bm{I}_N)^{-1}(\bm{\F}_e^{s})^{\top}\widehat{\X}_{n+1}^{*,s}(u_e),
\end{equation*}
where $\bm{I}_N$ is an $(N\times N)$ identity matrix. When $\lambda\rightarrow 0$, $\widehat{\bm{\beta}}_{n+1}^{s, \text{RR}}$ approaches $\widehat{\bm{\beta}}_{n+1}^{s, \text{OLS}}$; when $\lambda \rightarrow \infty$, $\widehat{\bm{\beta}}_{n+1}^{s, \text{RR}}\rightarrow 0$; when $0<\lambda<\infty$, $\widehat{\bm{\beta}}_{n+1}^{s, \text{RR}}$ is a weighted average between 0 and $\widehat{\bm{\beta}}_{n+1}^{s, \text{OLS}}$. With an optimally selected $\lambda$ value, the RR forecast of $\X_{n+1}^s(u_l)$ can be given by
\begin{equation*}
\widehat{\X}_{n+1}^{s, \text{RR}}(u_l) = \widehat{\mu}^{s}(u_l) + \sum^N_{i=1}\widehat{\beta}_{n+1,i}^{s, \text{RR}}\widehat{\phi}_i^s(u_l).
\end{equation*}

\subsubsection{Penalized least squares (PLS)}

As an alternative shrinkage estimator, the PLS coefficient estimate minimises a penalised residual sum of squares:
\begin{equation*}
\argmin_{\bm{\beta}^s_{n+1}}\left\{(\widehat{\X}_{n+1}^{*,s}(u_e) - \bm{\F}^s_e\bm{\beta}^s_{n+1})^{\top}(\widehat{\X}_{n+1}^{*,s}(u_e) - \bm{\F}^s_e\bm{\beta}^s_{n+1})+\lambda (\bm{\beta}^s_{n+1} - \widehat{\bm{\beta}}_{n+1|n}^{s, \text{TS}})^{\top}(\bm{\beta}^s_{n+1} - \widehat{\bm{\beta}}_{n+1|n}^{s, \text{TS}})\right\},
\end{equation*}
where $\widehat{\bm{\beta}}_{n+1|n}^{s, \text{TS}}$ denotes the one-step-ahead point forecasts of the principal component scores using a univariate time-series forecasting method, such as exponential smoothing of \cite{HKO+08}. Instead of shrinking towards zero, the PLS regression coefficient estimates shrink towards the TS coefficient estimates. By taking the first-order derivative with respect to $\bm{\beta}_{n+1}$, we obtain
\begin{equation*}
\widehat{\bm{\beta}}_{n+1}^{s, \text{PLS}} = \left[(\bm{\F}_e^s)^{\top}\bm{\F}^s_e+\lambda\bm{I}_N\right]^{-1}\left[(\bm{\F}_e^s)^{\top}\widehat{\X}_{n+1}^{*,s}(u_e)+\lambda \widehat{\bm{\beta}}_{n+1|n}^{s, \text{TS}}\right].
\end{equation*}
When the shrinkage parameter $\lambda\rightarrow 0$, $\widehat{\bm{\beta}}_{n+1}^{s, \text{PLS}}$ approaches $\widehat{\bm{\beta}}_{n+1}^{s, \text{OLS}}$; when $\lambda\rightarrow \infty$, $\widehat{\bm{\beta}}_{n+1}^{s, \text{PLS}}$ approaches $\widehat{\bm{\beta}}_{n+1}^{s, \text{TS}}$; when $0<\lambda<\infty$, $\widehat{\bm{\beta}}_{n+1}^{s, \text{PLS}}$ is a weighted average between $\widehat{\bm{\beta}}_{n+1}^{s, \text{OLS}}$ and $\widehat{\bm{\beta}}_{n+1}^{s, \text{TS}}$. With an optimally selected $\lambda$ value, the PLS forecast of $\X_{n+1}^s(u_l)$ can be given by
\begin{equation*}
\widehat{\X}_{n+1}^{s, \text{PLS}}(u_l) = \widehat{\mu}^s(u_l) + \sum^N_{i=1}\widehat{\beta}_{n+1,i}^{s, \text{PLS}}\widehat{\phi}_i^s(u_l).
\end{equation*}

\subsubsection{Selection of shrinkage parameters}

\begin{figure}[!htb]
\begin{center}
\begin{tikzpicture}[box/.style = {rectangle, draw=black, thick, minimum width=3cm, minimum height=1.2cm, align=center},  ->, thick]
\node[box] (train) {Training data};
\node[box, right=1.5cm of train] (val) {Validation data};
\node[box, right=1.5cm of val] (test) {Testing data};

\draw[->] (train) -- (val);
\draw[->] (val) -- (test);
\end{tikzpicture}
\end{center}
\caption{Illustration of the sample splitting. To select the optimal tuning parameter, a model is constructed using data from the training set to forecast data in the validation set. To evaluate forecast accuracy, a model is constructed using data from the training and validation sets to forecast data in the testing set. The validation dataset is used to select the optimal tuning parameters.}\label{fig:2}
\end{figure}
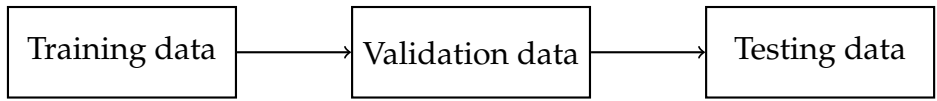

The superior performance of the RR and PLS estimators is based on an accurate selection of the $\lambda$ parameter. Depending on the updating period, the value of $\lambda$ should also change, particularly as more data points are observed throughout the day. We split our data sample, consisting of each day of a week, into a training sample comprising curves 1 to 306 and a testing sample comprising curves 307 to 408. We further divide the training sample into the training set, consisting of curves 1 to 204, and the validation set, consisting of curves 205 to 306. Within the validation set, the optimal values of $\lambda$ for different updating periods were determined by minimising a forecast error. In Figure~\ref{fig:2}, we present a concept diagram.

\section{Interval forecast methods}\label{sec:5}

Prediction intervals are a valuable tool for measuring the probabilistic uncertainty associated with point forecasts. As emphasised in \cite{Chatfield93, Chatfield00}, it is important to provide interval forecasts as well as point forecasts to 
\begin{inparaenum}[(i)]
\item assess future uncertainty;
\item enable different strategies to be planned for a range of possible outcomes indicated by the interval forecasts;
\item compare forecasts from different methods more thoroughly; and
\item explore different scenarios based on different assumptions.
\end{inparaenum}

\subsection{Prediction interval calibration}\label{sec:5.1}

We split the data sample, consisting of 408 curves, into training, validation, and testing sets. Using the data in the training sample, we implement an expanding window forecast scheme to obtain the one-step-ahead forecast in the validation set. We have 102 forecasts compared to the holdout sample in the validation set. From these residual functions, we compute the pointwise standard deviation, denoted by
\begin{equation}
\delta^s(u) = \text{sd}\left[\widehat{\xi}_{1}^s(u), \widehat{\xi}_{2}^s(u), \dots,\widehat{\xi}_{102}^s(u)\right],\label{eq:sd}
\end{equation}
where $\widehat{\xi}^s_{\omega}(u) = P_{\omega}^s(u) - \widehat{P}_{\omega}^s(u)$, for $\omega=1, 2,\dots,102$. For a level of significance $\alpha$, customarily $\alpha=0.2$ or 0.05, we aim to determine $(\underline{\theta}^s_{\alpha}, \overline{\theta}^s_{\alpha})$ so that $(1-\alpha)\times 100\%$ of the residuals satisfy
\begin{equation*}
\underline{\theta}^s_{\alpha}\delta^s(u)\leq \widehat{\xi}^s_{\omega}(u)\leq \overline{\theta}^s_{\alpha}\delta^s(u).
\end{equation*}
For ease of optimisation, the constants $\underline{\theta}^s_{\alpha}$ and $\overline{\theta}^s_{\alpha}$ are chosen equal. By the law of large numbers, one may achieve
\begin{equation*}
pr[-\theta^s_{\alpha}\delta^s(u)\leq P^s_{n+1}(u) - \widehat{P}^s_{n+1}(u)\leq \theta^s_{\alpha}\delta^s(u)] \approx \frac{1}{102}\sum^{102}_{\omega=1}\mathds{1}\left[-\theta^s_{\alpha}\delta^s(u)\leq \widehat{\xi}^s_{\omega}(u)\leq \theta^s_{\alpha}\delta^s(u)\right],
\end{equation*}
where $\mathds{1}[\cdot]$ denotes the binary indicator function. For a given level of significance $\alpha$, the value of $\theta_{\alpha}$ is selected by achieving the smallest absolute difference between the empirical and nominal coverage probabilities \citep[see also][]{SH25}.

\subsection{Conformal prediction interval}\label{sec:5.2}

As an alternative, a popular methodology known as conformal prediction \citep{SV08} can be used to construct probabilistic forecasts calibrated on out-of-sample interval forecast errors. Conformal prediction is model-agnostic and provides a distribution-free method to construct prediction sets with a finite-sample coverage guarantee. From the absolute value of $\widehat{\xi}^s_{\omega}(u)$, we compute its $(1-\alpha)\times 100\%$ quantile, denoted by $q^s_{\alpha}(u)$. The prediction interval can be obtained as
\begin{equation}
\left[\widehat{P}_{n+1}^s(u) - q^s_{\alpha}(u), \widehat{P}_{n+1}^s(u) + q^s_{\alpha}(u)\right],\label{eq:conformal}
\end{equation}
where $\widehat{P}_{n+1}^s(u)$ denotes the one-day-ahead point forecasts for the data in the testing set. A limitation is that the conformal prediction approach has a fixed width of $2 \times q_{\alpha}^s(u)$, which is not dependent on $\widehat{\X}_{n+1}^s(u)$ \citep{YPC19}. However, without the selection of any tuning parameter, the approach of the conformal prediction interval is computationally advantageous and works particularly well when the sample sizes in the validation and testing sets are reasonably large \citep{DDR24}.

\subsection{Updating interval forecasts}\label{sec:5.3}

As new intraday data become available sequentially, prediction intervals can be dynamically updated to improve the accuracy of interval forecasts. Within the validation set, the optimal ridge penalty parameter $\lambda$ is selected based on point forecast errors. In addition, the associated residual functions are computed. Using newly observed intraday data, the residuals corresponding to the remaining forecast period generally tend to diminish in magnitude, facilitating the refinement of the prediction intervals. This refinement can be performed using either the sd approach in~\eqref{eq:sd} or the conformal prediction in~\eqref{eq:conformal}.

\section{Measures of point and interval forecast accuracy}\label{sec:6}

We evaluate the performance of the forecast methods using different measures: mean absolute percentage error (MAPE), coverage probability difference (CPD) and mean interval scores.

\subsection{Mean absolute percentage error}\label{sec:6.1}

We compute the point forecasts for the data in the testing set and evaluate the forecast accuracy by MAPE. MAPE measures the closeness of the forecasts compared to the actual values, regardless of the sign. It can be expressed as
\begin{equation*}
\text{MAPE}^s(u_j) = \frac{1}{102}\sum^{102}_{\omega=1}\frac{\left|P_{\omega}^s(u_j) - \widehat{P}_{\omega}^s(u_j)\right|}{P_{\omega}^s(u_j) }\times 100\%, \qquad j=1,\dots,p.
\end{equation*}
where $P_{\omega}^s(u_j)$ denotes the actual holdout sample for the $j$\textsuperscript{th} intraday period on day $\omega$ for a particle size $s$, and $\widehat{P}_{\omega}^s(u_j)$ denotes the one-day-ahead point forecasts for the holdout sample in the validation or testing set, in which each of them contains 102 curves.

\subsection{Coverage probability difference and mean interval scores}\label{sec:6.2}

To evaluate the interval forecast accuracy, we utilise the CPD, which is the absolute difference between the empirical and nominal coverage probabilities. For each year in the validation or testing set, the one-step-ahead prediction intervals were computed at the $(1-\alpha)\times 100\%$ nominal coverage probability. Let $[\widehat{P}_{\omega}^{s,\text{lb}}(u_j), \widehat{P}_{\omega}^{s,\text{ub}}(u_j)]$ denote the lower and upper bounds, respectively, for $\omega$\textsuperscript{th} in the testing set. For a given particle size $s$ at a particular hour of the day, the empirical coverage probability is defined as
\begin{equation*}
\text{ECP}^s(u_j) = 1 - \frac{1}{102}\sum^{102}_{\omega=1} \left[\mathds{1}(P_{\omega}^s(u_j)>\widehat{P}_{\omega}^{s,\text{ub}}(u_j)) + \mathds{1}(P_{\omega}^s(u_j)<\widehat{P}_{\omega}^{s,\text{lb}}(u_j))\right],
\end{equation*}
where $\mathds{1}(\cdot)$ denotes the binary indicator function. The CPD is the absolute difference between the empirical and nominal coverage probabilities.

As defined by \cite{GR07} and \cite{GK14}, a scoring rule for the interval forecast at the time point $u_j$ is
\begin{align*}
S_{\alpha}[\widehat{P}_{\omega}^{s,\text{lb}}(u_j), \widehat{P}_{\omega}^{s,\text{ub}}(u_j), P_{\omega}^s(u_j)] = [\widehat{P}_{\omega}^{s,\text{ub}}(u_j)-\widehat{P}_{\omega}^{s,\text{lb}}(u_j)] &+\frac{2}{\alpha}[\widehat{P}_{\omega}^{s,\text{lb}}(u_j)-P^s_{\omega}(u_j)]\mathds{1}[P_{\omega}^s(u_j)<\widehat{P}_{\omega}^{s,\text{lb}}(u_j)]\\
&+\frac{2}{\alpha}[P^s_{\omega}(u_j) - \widehat{P}^{s,\text{ub}}_{\omega}(u_j)]\mathds{1}[P_{\omega}^s(u_j)>\widehat{P}_{\omega}^{s,\text{ub}}(u_j)],
\end{align*}
where $\alpha$ denotes a level of significance. Averaging over 102 days, the mean interval score is given by
\begin{equation*}
\overline{S}_{\alpha}^s(u_j)=\frac{1}{102}\sum^{102}_{\omega=1}S_{\alpha}\big[\widehat{P}_{\omega}^{s,\text{lb}}(u_j), \widehat{P}_{\omega}^{s,\text{ub}}(u_j), P_{\omega}^s(u_j)\big].
\end{equation*}
The mean interval score rewards a narrow prediction interval, provided that the actual observations lie within the prediction interval.

\section{Results}\label{sec:7}

\subsection{Expanding window scheme}\label{sec:7.1}

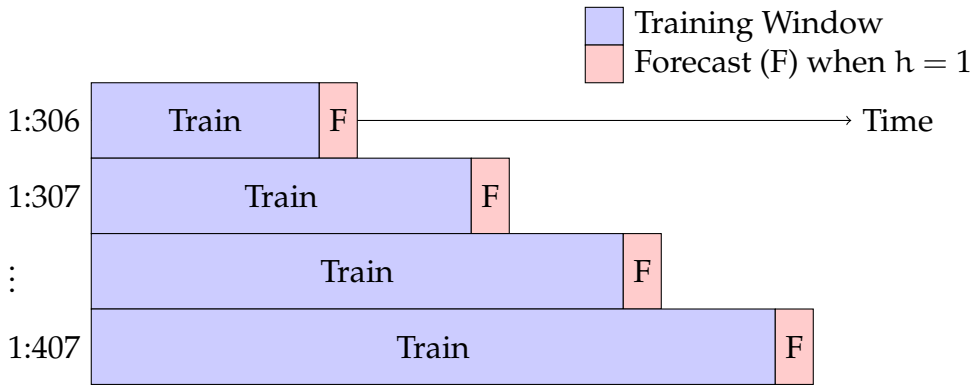
\begin{figure}[!htb]
\begin{center}
\begin{tikzpicture}
\draw[->] (0,0) -- (10,0) node[right] {Time};
    
\draw[fill=blue!20] (0,-0.5) rectangle (3,0.5) node[midway] {Train};
\draw[fill=red!20] (3,-0.5) rectangle (3.5,0.5) node[midway] {F};
    
\draw[fill=blue!20] (0,-1.5) rectangle (5,-0.5) node[midway] {Train};
\draw[fill=red!20] (5,-1.5) rectangle (5.5,-0.5) node[midway] {F};
    
\draw[fill=blue!20] (0,-2.5) rectangle (7,-1.5) node[midway] {Train};
\draw[fill=red!20] (7,-2.5) rectangle (7.5,-1.5) node[midway] {F};
    
\draw[fill=blue!20] (0,-3.5) rectangle (9,-2.5) node[midway] {Train};
\draw[fill=red!20] (9,-3.5) rectangle (9.5,-2.5) node[midway] {F};
    
\node[left] at (0,0) {1:306};
\node[left] at (0,-1) {1:307};
\node[left] at (0,-2) {\hspace{-0.8in}{$\vdots$}};
\node[left] at (0,-3) {1:407};
    
\draw[fill=blue!20] (6.5,1) rectangle (7,1.5);
\node[right] at (7,1.25) {Training Window};
\draw[fill=red!20] (6.5,0.5) rectangle (7,1);
\node[right] at (7,0.75) {Forecast (F) when $h=1$};
\end{tikzpicture}
\end{center}
\caption{A diagram of the expanding-window one-step-ahead forecast scheme.}\label{fig:3}
\end{figure}

An expanding window scheme is implemented to evaluate model and parameter stability over time. The expanding window analysis determines the variability of the model's parameters by iteratively computing parameter estimates and their resultant forecasts over an expanding window. Using the first 306 curves, we produce an one-day-ahead forecast for the 307\textsuperscript{th} curve. Then, we increase the training sample by one day. Using the first 307 curves, we produce an one-day-ahead forecast for the 308\textsuperscript{th} curve. This procedure continues until the training sample encompasses the entire dataset. In doing so, there are 102 days in the testing sample for evaluation of the forecast accuracy. Similarly, we can divide the 306 curves further into a training set and a 102-day validation set for parameter selection. In Figure~\ref{fig:3}, we present a concept diagram.

%There are two strategies to implement the expanding-window forecast scheme: First, we can sort the data according to the day of the week, and there are 408 observations for each day. For instance, we consider the historical data on Mondays to produce the one-day-ahead forecast for Monday. In doing so, we may have a homogeneous group and term this strategy as a weekday cluster approach. Second, we utilise all data to produce the forecasts and sort the forecasts according to the day of the week. In this strategy, a rationale is that the one-day-ahead forecasts strongly depend on the values on the current day, regardless of the day of the week.

%\vspace{-.2in}

\subsection{Evaluation of point forecast accuracy}\label{sec:7.2}

For producing one-day-ahead forecasts, we consider two approaches: the MLFTS method and a combination of the functional factor model and the MLFTS method, as described in Section \ref{sec:3}. In the second approach, the functional factor model can remove the common trend and residuals can then be modelled via the MLFTS method. Via the expanding-window forecast scheme, we sort the data according to the day of the week, and there are 408 observations for each day (number of weeks in the dataset). For example, we consider historical data on Mondays to produce the one-day-ahead forecast for Monday. In doing so, we may have a homogeneous group. The temporal dependence is generally stronger for the same day between two successive weeks rather than two successive weekdays when producing one-step-ahead point forecasts.
\begin{table}[!htb]
\centering
\tabcolsep 0.15in
\caption{As measured by the MAPE, a comparison of point forecast accuracy of the functional time-series methods, averaged over 24 hours and 51 particle sizes. We consider three ways of segmenting a univariate time series, namely, each weekday from historical weeks (Weekday), every consecutive day (Day), and every consecutive week (Week).}\label{tab:1}
\begin{tabular}{@{}llrrrrrrr@{}}
\toprule
Data & Method & Mon & Tue & Wed & Thu & Fri & Sat & Sun \\
\midrule 
Weekday 	& MLFTS 			& 100.89 	& 86.84 & 99.55 & 99.68 & 92.52 & 83.72 & 109.25 \\
		& Factor + MLFTS 	& 99.39 	& 85.76 & 97.49 & 98.47 & 92.59 & 84.76 & 109.06 \\
\midrule        
Day     & MLFTS & 88.79 & 75.95 & 89.96 & 62.05 & 67.09 & 80.00 & 113.81 \\ 
        & Factor + MLFTS & 89.33 & 76.31 & 89.52 & 61.60 & 67.86 & 80.95 & 113.83 \\ 
\midrule        
Week    & MLFTS & 77.91 & 70.12 & 81.64 & 81.82 & 81.61 & 68.01 & 87.23 \\ 
        & Factor + MLFTS & 78.22 & 70.21 & 81.54 & 82.18 & 82.90 & 68.08 & 87.27 \\         
\bottomrule   
\end{tabular}
\end{table}

In Table~\ref{tab:1}, we compare the one-step-ahead point forecast accuracy of two functional time series methods. The first method follows the MLFTS framework, which decomposes multiple functional time series into a common trend and series-specific residual trends. The second method builds upon the MLFTS approach, further incorporating a functional factor model to remove the trend component. There are three ways to form our functional data: 
\begin{inparaenum}
\item[1)] We segment our data into each day of the week for 408 weeks (i.e., Weekday);
\item[2)] We consider $408\times 7=2856$ consecutive days, which consist of weekdays and weekends, as our functional data;
\item[3)] We consider the weekly curve as our functional data for 408 weeks.
\end{inparaenum}
Among the three ways to form our functional data, the weekly curves provide the smallest point forecast errors for most days, except for Thursday and Friday. The second way is computationally time-consuming for training models using the expanding-window approach, and thus is not feasible for the purpose of dynamic updating. Using the \Verb|dm.test| function in the forecast package \citep{HAB+25}, with equal performance superiority, we compute the $p$-value for the functional data constructed from each weekday and week and obtain the $p$-value of zero. For the weekly curves, the $p$-value of the \citeauthor{DM95}'s \citeyearpar{DM95} test is given as 0.1481 between the MLFTS and Factor+MLFTS methods, and this indicates equal forecast superiority.

\begin{figure}[!htb]
\centering
\subfloat[Averaged over 51 sizes]
{\includegraphics[width=8.78cm]{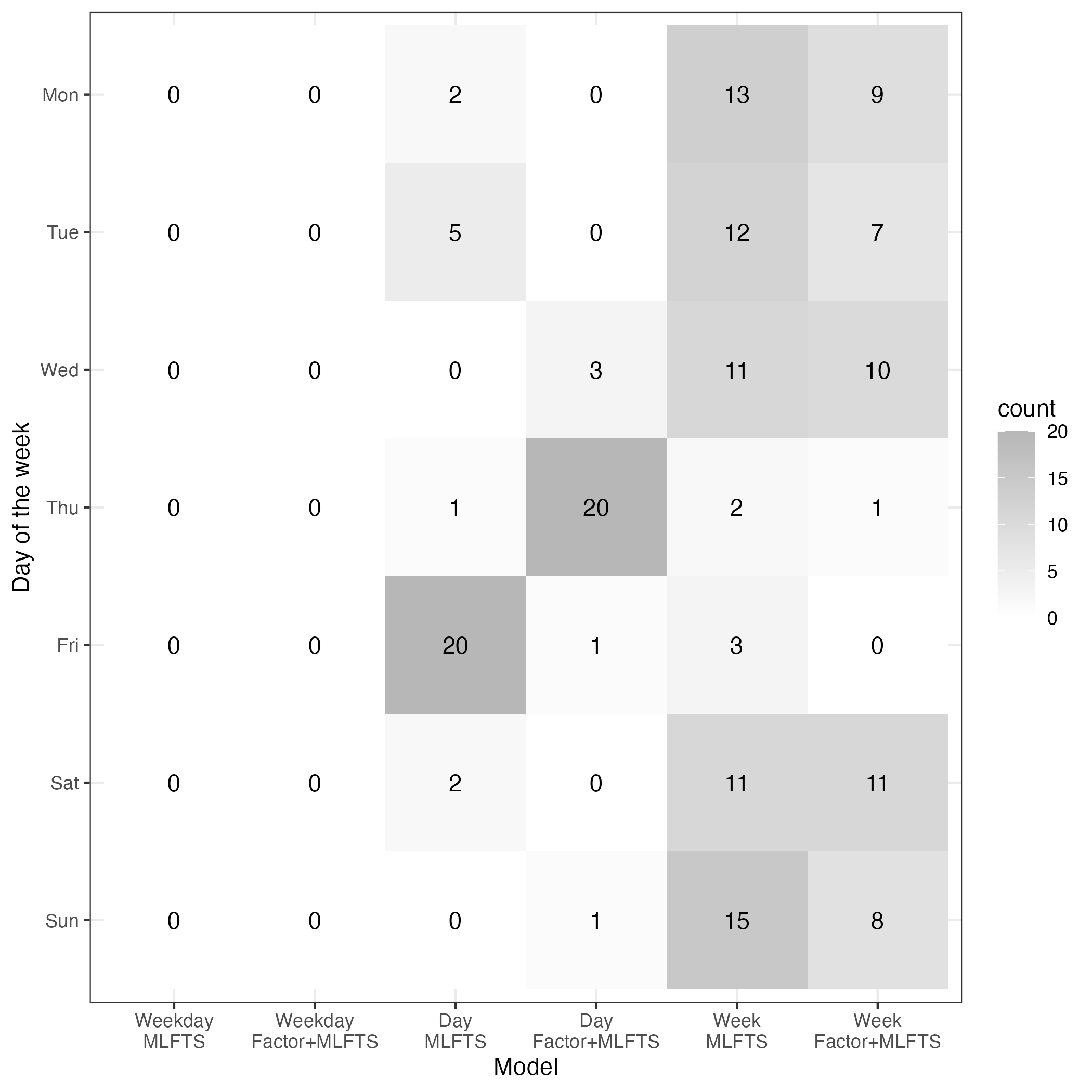}\label{fig:4a}}
\quad
\subfloat[Averaged over 24 hours]
{\includegraphics[width=8.78cm]{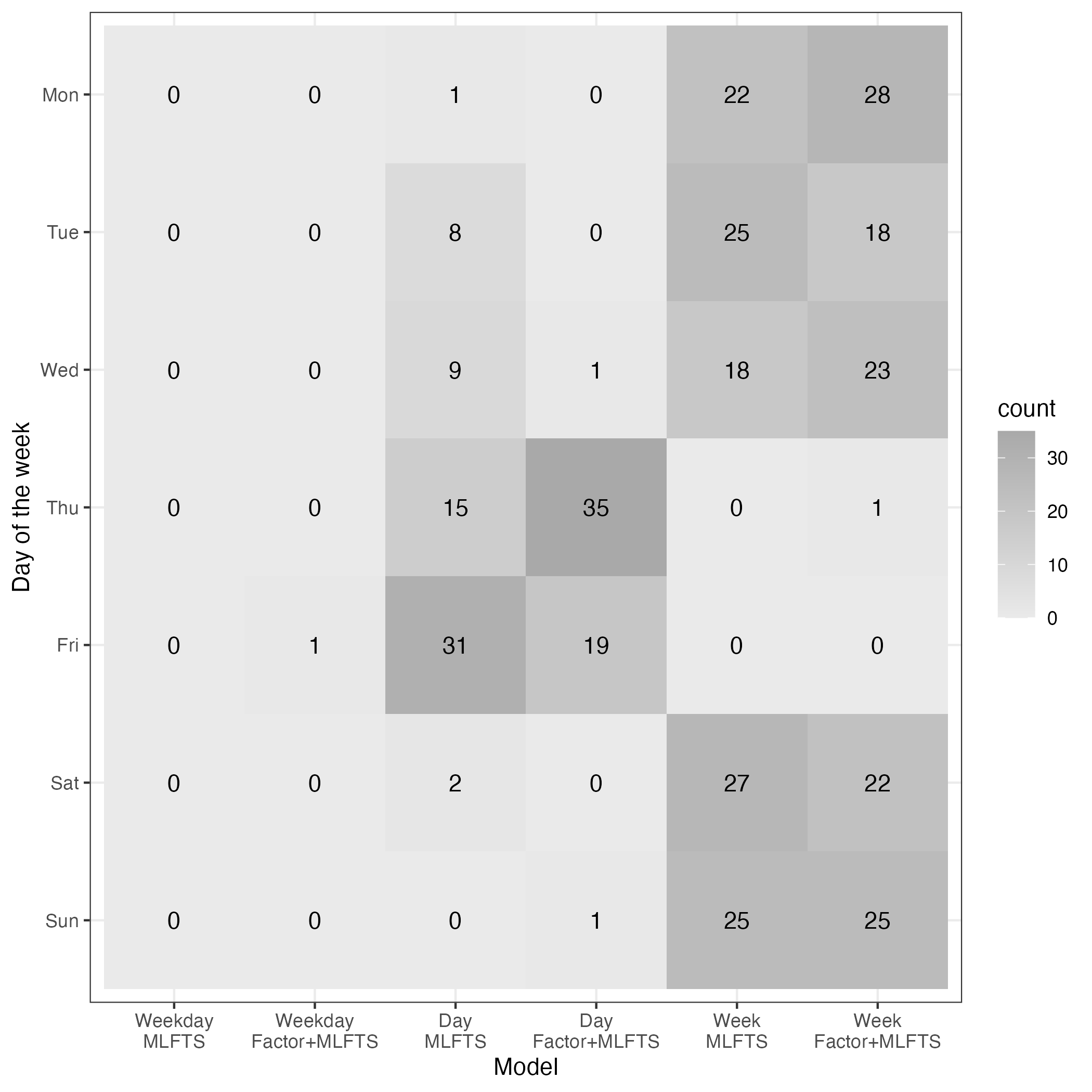}\label{fig:4b}}
\caption{Frequency of the most accurate functional time-series forecasting method for each day of the week. In (a), each row sums to 24 hours in a day. In (b), each row sums to 51 particulate matter sizes. We consider three ways of segmenting a univariate time series, namely each weekday from historical weeks (Weekday), every consecutive day (Day) and every consecutive week (Week).}\label{fig:4}
\end{figure}

In Figure~\ref{fig:4a}, we present a heatmap with a grey colour scaling. For the weekly curves, we segment its forecasts according to each day of the week, to facilitate a fair comparison. For each day of the week, we count the number of times a model provides the smallest MAPE averaged over 51 particle sizes. Each row of the heatmap sums up to 24 hours a day. Similarly, we also display another heatmap in Figure~\ref{fig:4b} where we count the number of times a model provides the smallest MAPE averaged over 24 hours. Each row of the heatmap sums up to 51 particle sizes. Between the two models, it is advantageous to consider the combination of the factor model and MLFTS for most days, except Saturday.

\begin{figure}[!htb]
\centering
\subfloat[Averaged over 51 sizes]
{\includegraphics[width=8.78cm]{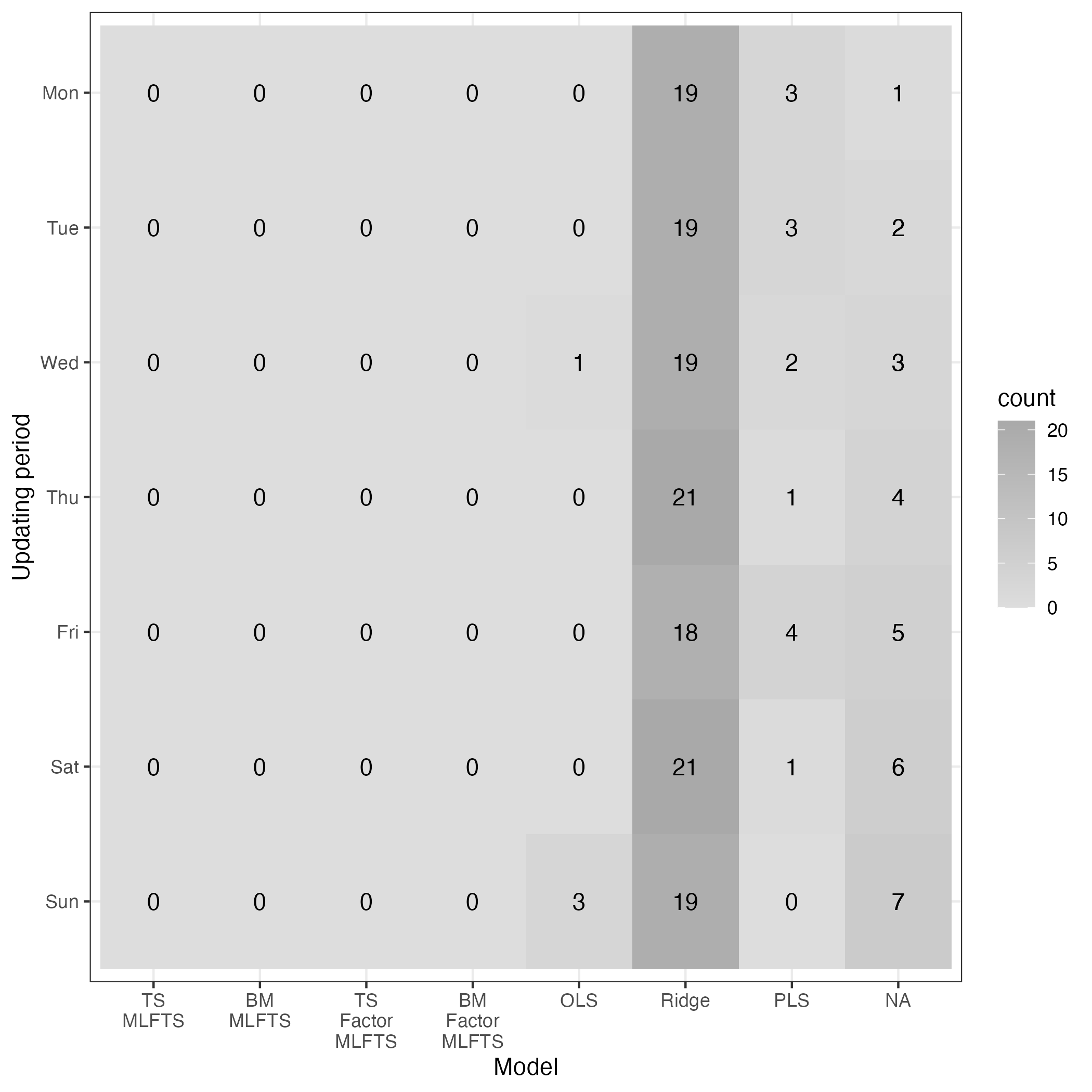}\label{fig:6a}}
\quad
\subfloat[Averaged over 24 hours]
{\includegraphics[width=8.78cm]{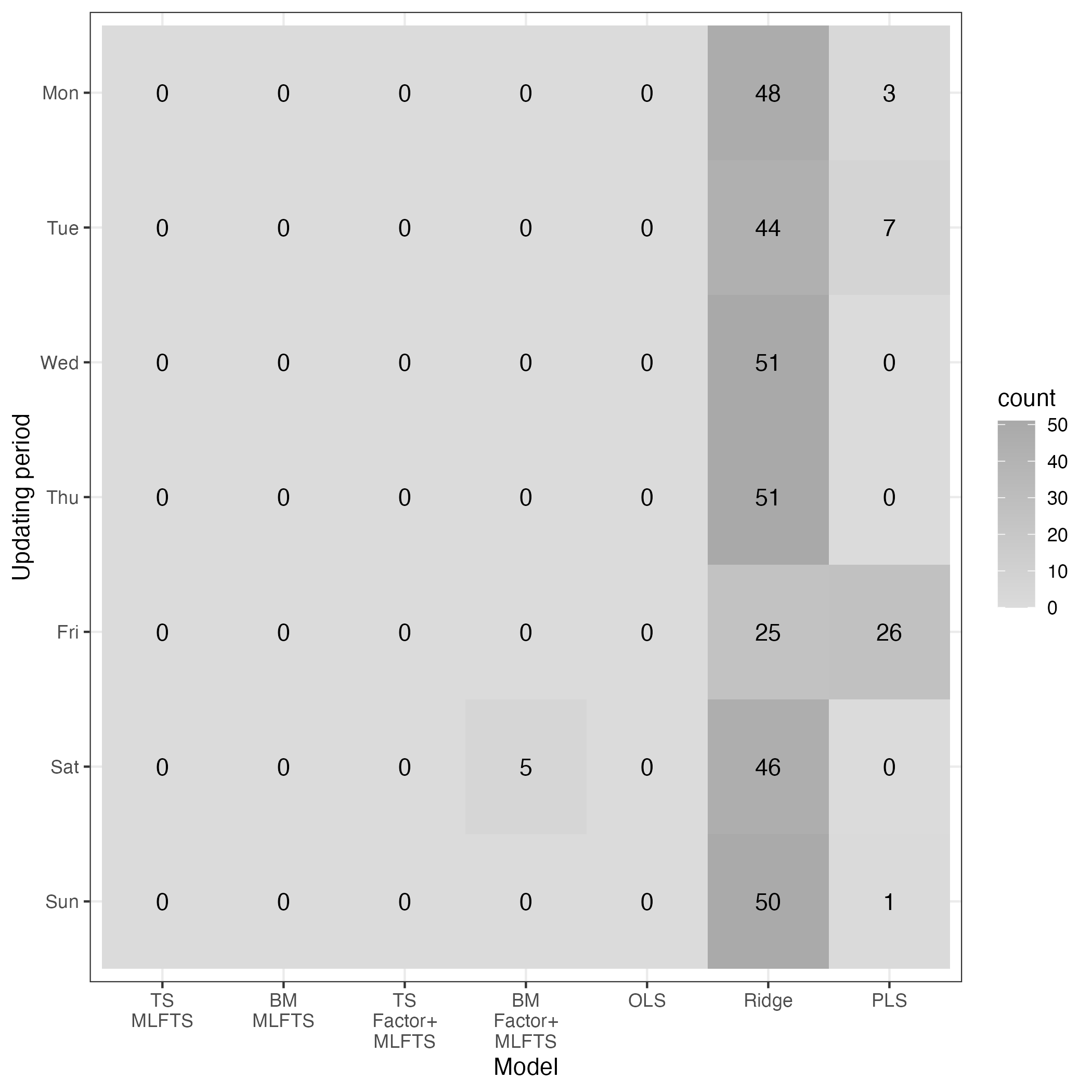}\label{fig:6b}}
\caption{Frequency of the most accurate dynamic updating method for each day of the week. In (a), each row sums to 24 hours in a day. In (b), each row sums to 51 particulate matter sizes.}\label{fig:6}
\end{figure}

In Figure~\ref{fig:6}, we consider the block-moving approach and regression-based approaches based on the OLS, ridge, and PLS estimators. In contrast to the functional time-series forecasting method, implementing a dynamic updating approach is advantageous in achieving better forecast accuracy. Between the two dynamic updating approaches, the regression-based approach shows superior forecast accuracy. Among the regression-based approach, the ridge estimator provides the most accurate forecasts.

\begin{figure}[!htb]
\centering
\includegraphics[width=8.74cm]{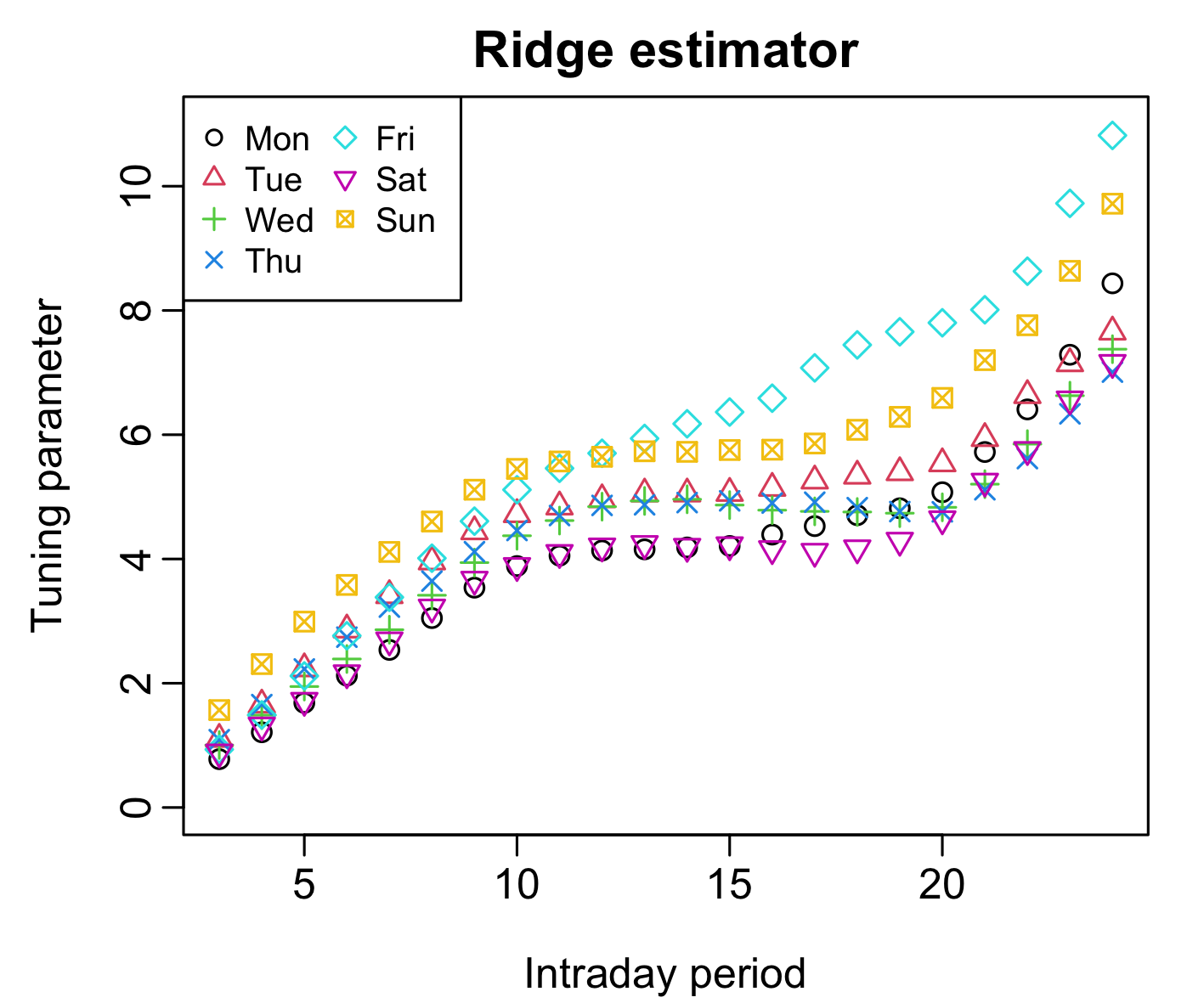}
\quad
\includegraphics[width=8.74cm]{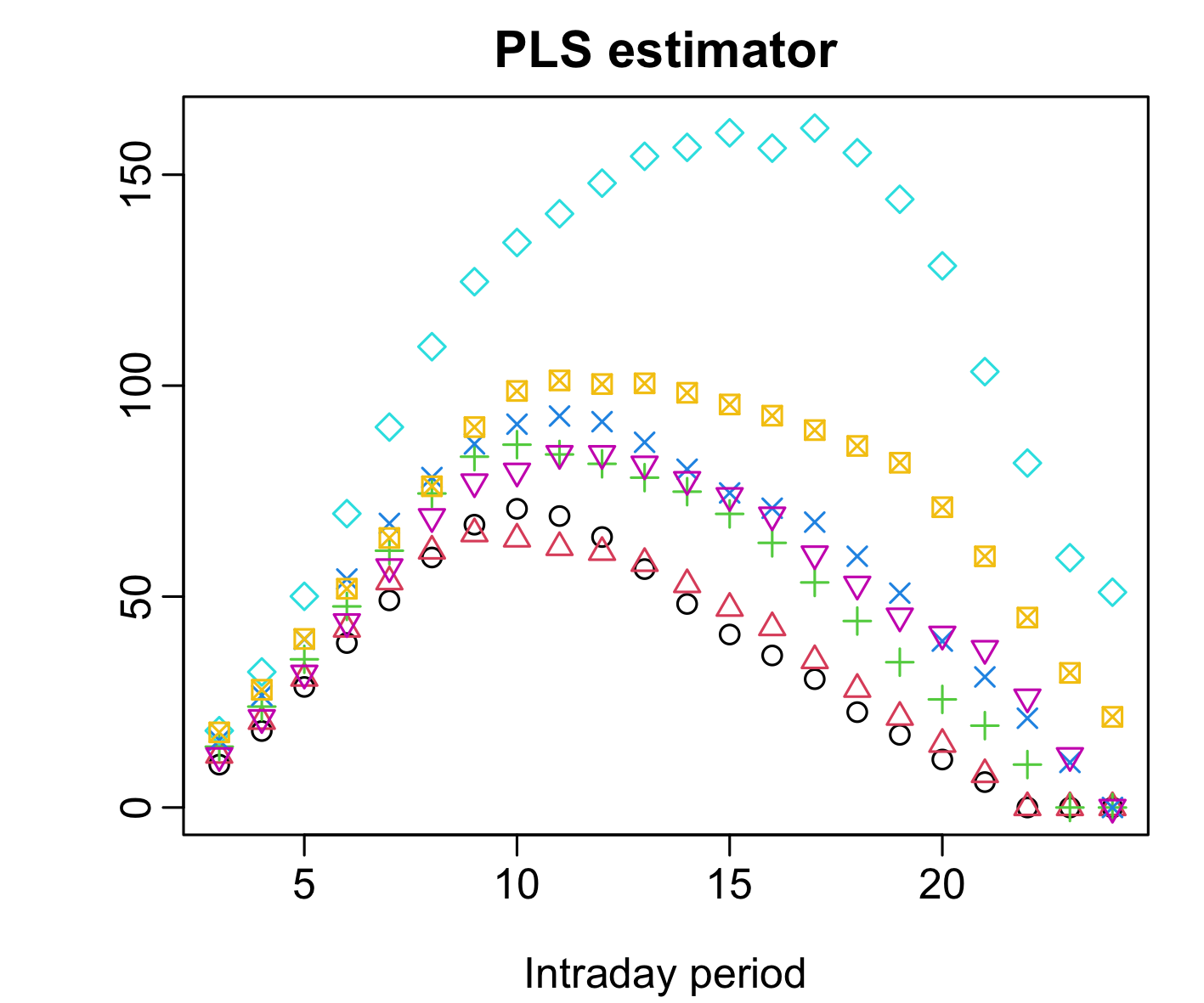}
\caption{Optimally selected tuning parameter $\lambda$ based on the MAPE for different updating periods for each day of the week.}\label{fig:5}
\end{figure}

In Figure~\ref{fig:5}, we plot the selected optimal (non-negative) tuning parameters in the ridge and PLS estimators throughout the intraday periods. These tuning parameters are chosen on the basis of the smallest overall MAPE in the validation data. Although a general pattern appears to exist, it also exhibits daily variation. For the ridge estimator, the estimated regression coefficient gradually shrinks towards zero; for the PLS estimator, the estimated regression coefficient gradually shrinks towards the OLS estimator throughout the day. Regarding the PLS estimator, the tuning parameters selected for Fridays tend to be larger than other days. Although it is difficult to pinpoint an affirmative reasoning, we believe that it might be due to the fact that Fridays often act as a transition day, where people living near the measurement station in London spend weekends elsewhere, thus influencing the PM measurement \citep{BCW+09}.

\subsection{Evaluation of interval forecast accuracy}\label{sec:7.3}

In Table~\ref{tab:4}, we compare the interval forecast accuracy based on the CPD and the mean interval scores at the nominal coverage probabilities of 80\% and 95\%. At the 80\% nominal coverage probability, the differences between the two prediction interval construction methods are marginal. However, at the 95\% nominal coverage probability, the standard deviation approach shows a slight advantage over the conformal prediction. Based on the CPD, the \textit{factor + MLFTS} method provides a smaller difference than the ones obtained from the MLFTS method. Oftentimes, the superiority in CPD is attributed to the larger interval scores. 
\begin{center}
\tabcolsep 0.042in
\renewcommand{\arraystretch}{0.9}
\begin{longtable}{@{}lllcccccccc@{}}
\caption{A comparison of the interval forecast accuracy, as measured by the CPD and mean interval scores, at the nominal coverage probabilities of 80\% and 95\% between the MLFTS and factor + MLFTS methods. Further, we consider two ways of constructing our functional data, constructed either from 24 or 168 hourly data points.}\label{tab:4} \\
\toprule
& & & \multicolumn{4}{c}{CPD} & \multicolumn{4}{c}{Mean interval score}  \\
\cmidrule(lr){4-7}\cmidrule(lr){8-11}
& &	& \multicolumn{2}{c}{MLFTS} & \multicolumn{2}{c}{Factor + MLFTS} & \multicolumn{2}{c}{MLFTS} & \multicolumn{2}{c}{Factor + MLFTS} \\
\cmidrule(lr){4-5}\cmidrule(lr){6-7}\cmidrule(lr){8-9}\cmidrule(lr){10-11}
Method & $\alpha$ & Day & sd & conformal & sd & conformal & sd & conformal & sd & conformal  \\
\midrule
\endfirsthead
\toprule
& & & \multicolumn{4}{c}{CPD} & \multicolumn{4}{c}{Mean interval score}  \\
\cmidrule(lr){4-7}\cmidrule(lr){8-11}
& &	& \multicolumn{2}{c}{MLFTS} & \multicolumn{2}{c}{Factor + MLFTS} & \multicolumn{2}{c}{MLFTS} & \multicolumn{2}{c}{Factor + MLFTS} \\
\cmidrule(lr){4-5}\cmidrule(lr){6-7}\cmidrule(lr){8-9}\cmidrule(lr){10-11}
Method & $\alpha$ & Day & sd & conformal & sd & conformal & sd & conformal & sd & conformal  \\
\midrule
\endhead
\multicolumn{11}{r}{{Continued on next page}} \\
\endfoot
\endlastfoot
Weekday & 0.2 &  Mon 	& 0.0191 & 0.0162 & 0.0183 & 0.0163 & 8982 & 8905 & 9106 & 9028 \\
	& &  Tue 	& 0.0324 & 0.0319 & 0.0272 & 0.0251 & 8754 & 8715 & 8804 & 8769 \\
	& &  Wed 	& 0.0308 & 0.0360 & 0.0306 & 0.0377 & 10049 & 9957 & 10204 & 10118 \\
	& &  Thu 	& 0.0241 & 0.0284 & 0.0218 & 0.0260 & 9115 & 9063 & 9249 & 9213 \\
	& &  Fri 	& 0.0202 & 0.0180 & 0.0149 & 0.0139 & 8645 & 8374 & 8806 & 8572 \\ 
	& &  Sat 	& 0.0398 & 0.0387 & 0.0369 & 0.0367 & 7604 & 7511 & 7668 & 7596 \\
	& &  Sun 	& 0.0233 & 0.0211 & 0.0169 & 0.0137 & 7929 & 7835 & 8021 & 7919 \\
\cmidrule{3-11}
	& & Mean	& 0.0271 & 0.0272 & 0.0238 & 0.0242 & 8726 & 8623 & 8837 & 8745 \\
\cmidrule{2-11}
& 0.05 &  Mon 	& 0.0201 & 0.0250 & 0.0194 & 0.0246 & 18241 & 18414 & 18411 & 18618 \\
	& &  Tue 	& 0.0085 & 0.0059 & 0.0081 & 0.0054 & 17552 & 17623 & 17634 & 17734 \\ 
	& &  Wed 	& 0.0177 & 0.0215 & 0.0182 & 0.0230 & 21046 & 21314 & 21358 & 21656 \\
	& &  Thu 	& 0.0301 & 0.0348 & 0.0291 & 0.0346 & 18319 & 18519 & 18523 & 18730 \\ 
	& &  Fri 	& 0.0179 & 0.0223 & 0.0163 & 0.0204 & 16610 & 16289 & 16673 & 16324 \\ 
	& &  Sat 	& 0.0115 & 0.0090 & 0.0118 & 0.0084 & 14225 & 14338 & 14312 & 14427 \\
	& &  Sun 	& 0.0057 & 0.0038 & 0.0047 & 0.0041 & 15637 & 15615 & 15673 & 15693 \\
\cmidrule{3-11}
	& & Mean 	& 0.0159 & 0.0175 & 0.0154 & 0.0172 & 17376 & 17445 & 17512 & 17597 \\
\midrule 
Week & 0.2 & Mon & 0.0154 & 0.0101 & 0.0159 & 0.0106 & 8722 & 8650 & 8805 & 8721 \\ 
          & & Tue & 0.0577 & 0.0243 & 0.0579 & 0.0186 & 8683 & 8522 & 8735 & 8575 \\ 
          & & Wed & 0.0073 & 0.0339 & 0.0047 & 0.0303 & 10051 & 9974 & 10114 & 10004 \\ 
          & & Thu & 0.0339 & 0.0131 & 0.0383 & 0.0156 & 8818 & 8771 & 8926 & 8877 \\ 
          & & Fri & 0.0068 & 0.0102 & 0.0078 & 0.0087 & 8479 & 8221 & 8515 & 8242 \\ 
          & & Sat & 0.0097 & 0.0110 & 0.0102 & 0.0130 & 7249 & 7163 & 7330 & 7267 \\ 
          & & Sun & 0.0421 & 0.0143 & 0.0411 & 0.0177 & 7709 & 7601 & 7782 & 7663 \\ 
          & & Mean & 0.0247 & 0.0167 & 0.0251 & 0.0163 & 8530 & 8414 & 8601 & 8479 \\ 
\cmidrule{3-11}          
    & 0.05 & Mon & 0.0071 & 0.0189 & 0.0076 & 0.0190 & 173523 & 17614 & 17468 & 17822 \\ 
        & & Tue & 0.0094 & 0.0039 & 0.0088 & 0.0033 & 17222 & 17336 & 17239 & 17384 \\ 
        & & Wed & 0.0125 & 0.0193 & 0.0121 & 0.0197 & 20605 & 20882 & 20763 & 21092 \\ 
        & & Thu & 0.0245 & 0.0283 & 0.0255 & 0.0303 & 17568 & 17736 & 17682 & 17846 \\ 
        & & Fri & 0.0084 & 0.0088 & 0.0075 & 0.0080 & 16352 & 15933 & 16362 & 15945 \\ 
        & & Sat & 0.0055 & 0.0034 & 0.0058 & 0.0035 & 13603 & 13595 & 13665 & 13633 \\ 
        & & Sun & 0.0117 & 0.0012 & 0.0110 & 0.0011 & 14981 & 14863 & 15051 & 14955 \\ 
        \cmidrule{3-11}
        & & Mean & 0.0113 & 0.0120 & 0.0112 & 0.0121 & 16812 & 16852 & 16890 & 16954 \\ 
\bottomrule
\end{longtable}
\end{center}

\vspace{-.3in}

In Table~\ref{tab:4}, we consider our functional data constructed from 24 hourly data points in each day of the week or constructed from 168 hourly data in a week. We observe that the construction of weekly curves provides the smallest CPD and mean interval score at the nominal coverage probabilities of 80\% and 95\%. Using the \Verb|dm.test| function, we compute the $p$-value between the two methods for two ways of constructing functional curves. Based on the CPD, we find that the $p$-values are all zero, implying a statistically significant difference in the interval forecast accuracy.

\begin{figure}[!htb]
\centering
\subfloat[CPD]
{\includegraphics[width=8.67cm]{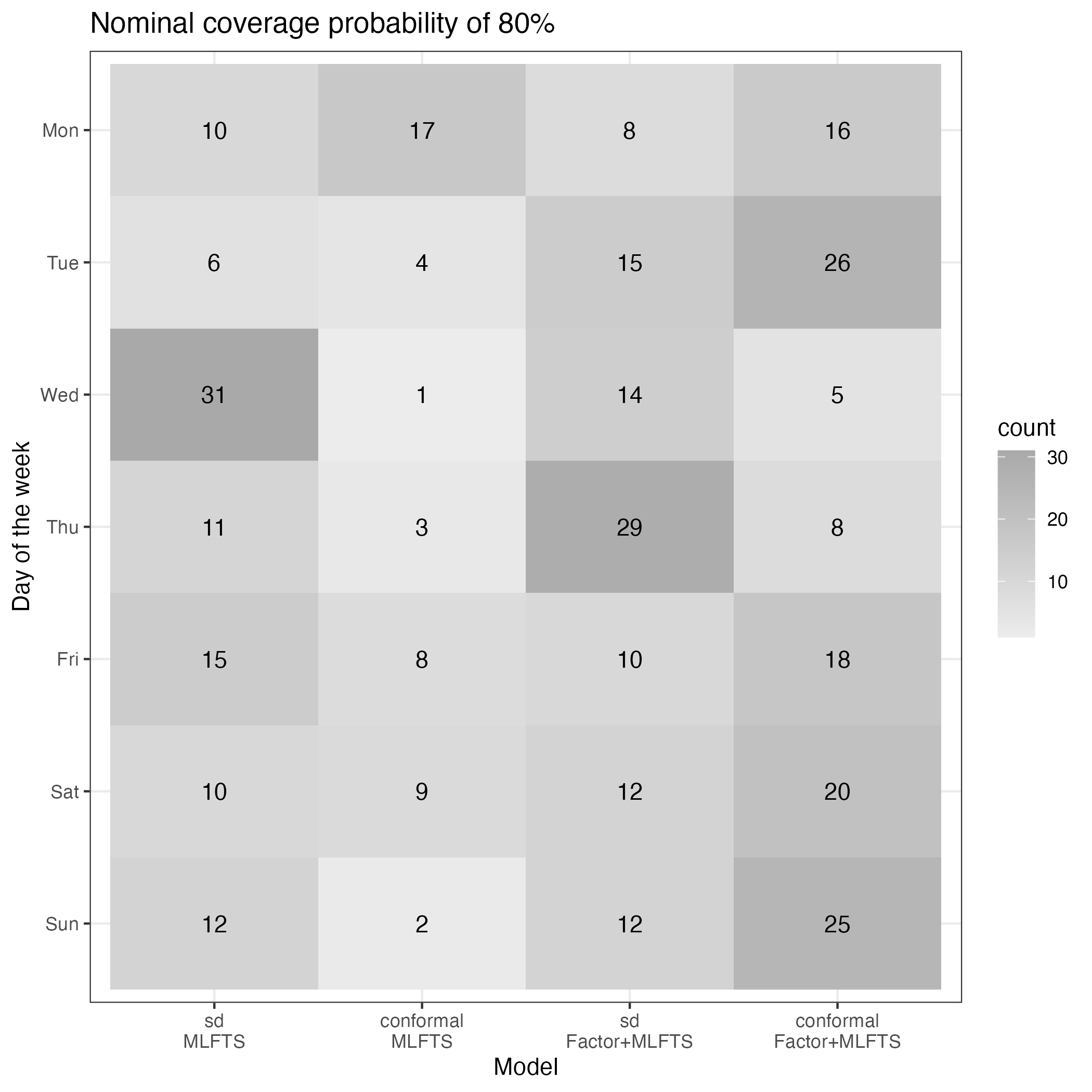}}
\quad
\subfloat[CPD]
{\includegraphics[width=8.67cm]{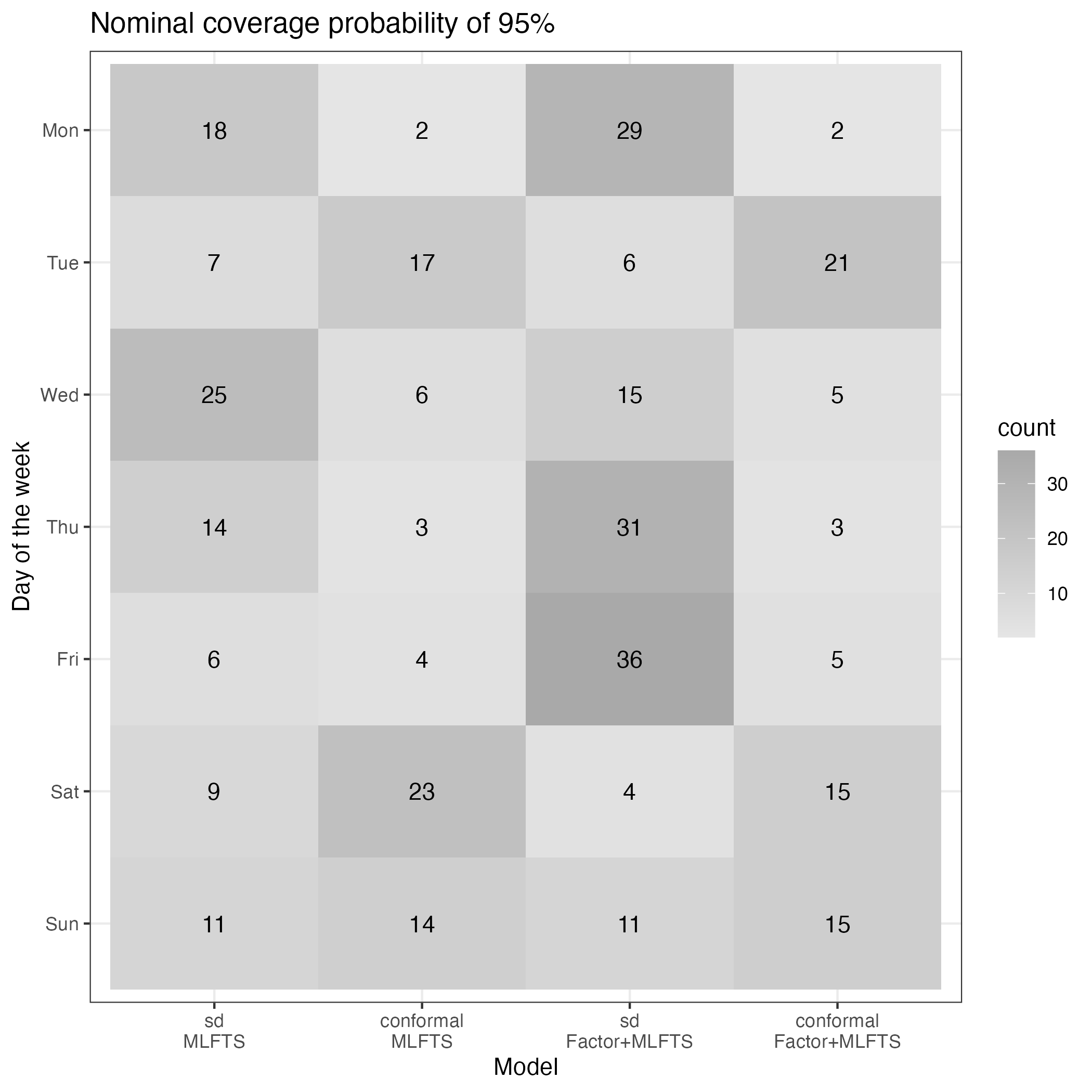}}
\\
\subfloat[Mean interval score]
{\includegraphics[width=8.67cm]{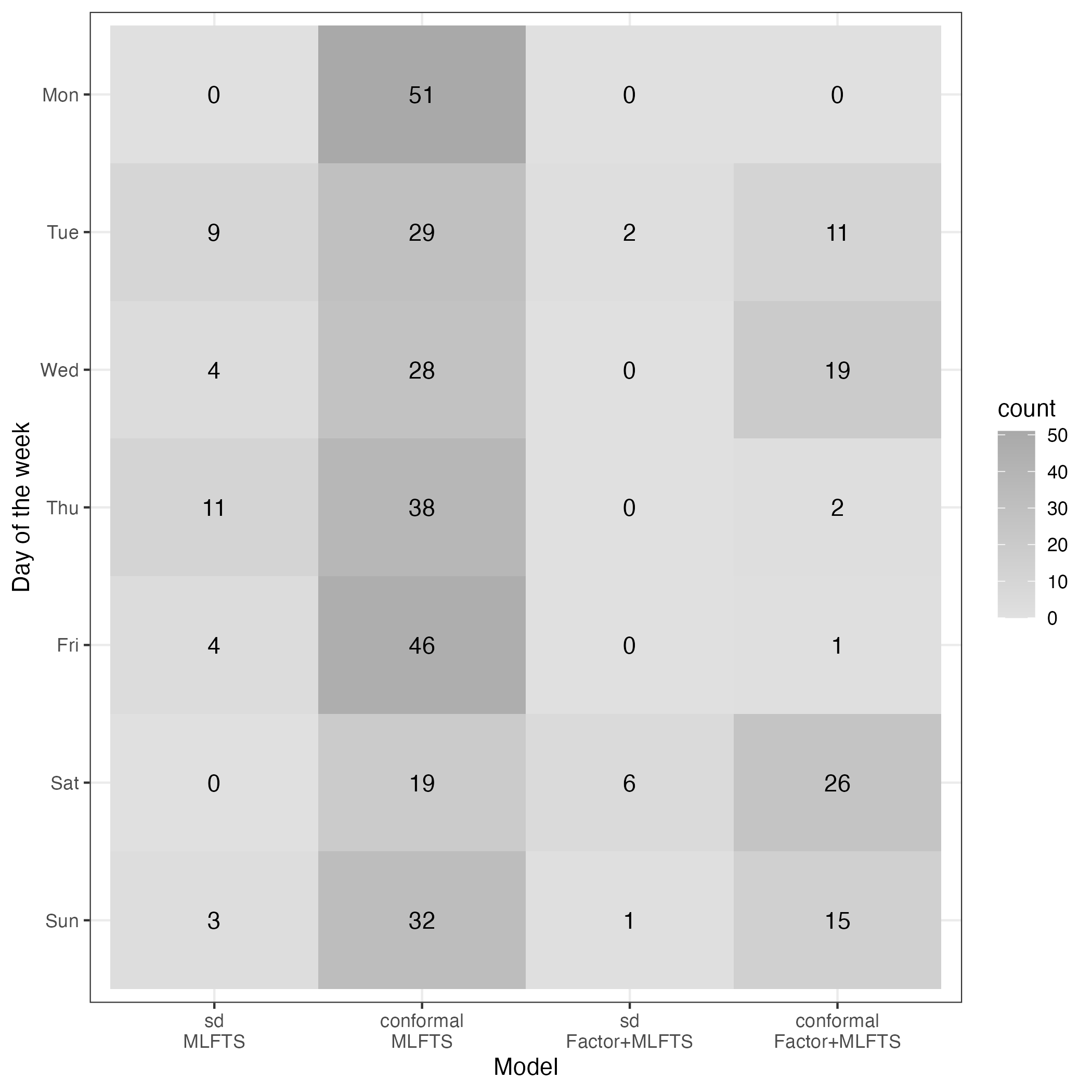}}
\quad
\subfloat[Mean interval score]
{\includegraphics[width=8.67cm]{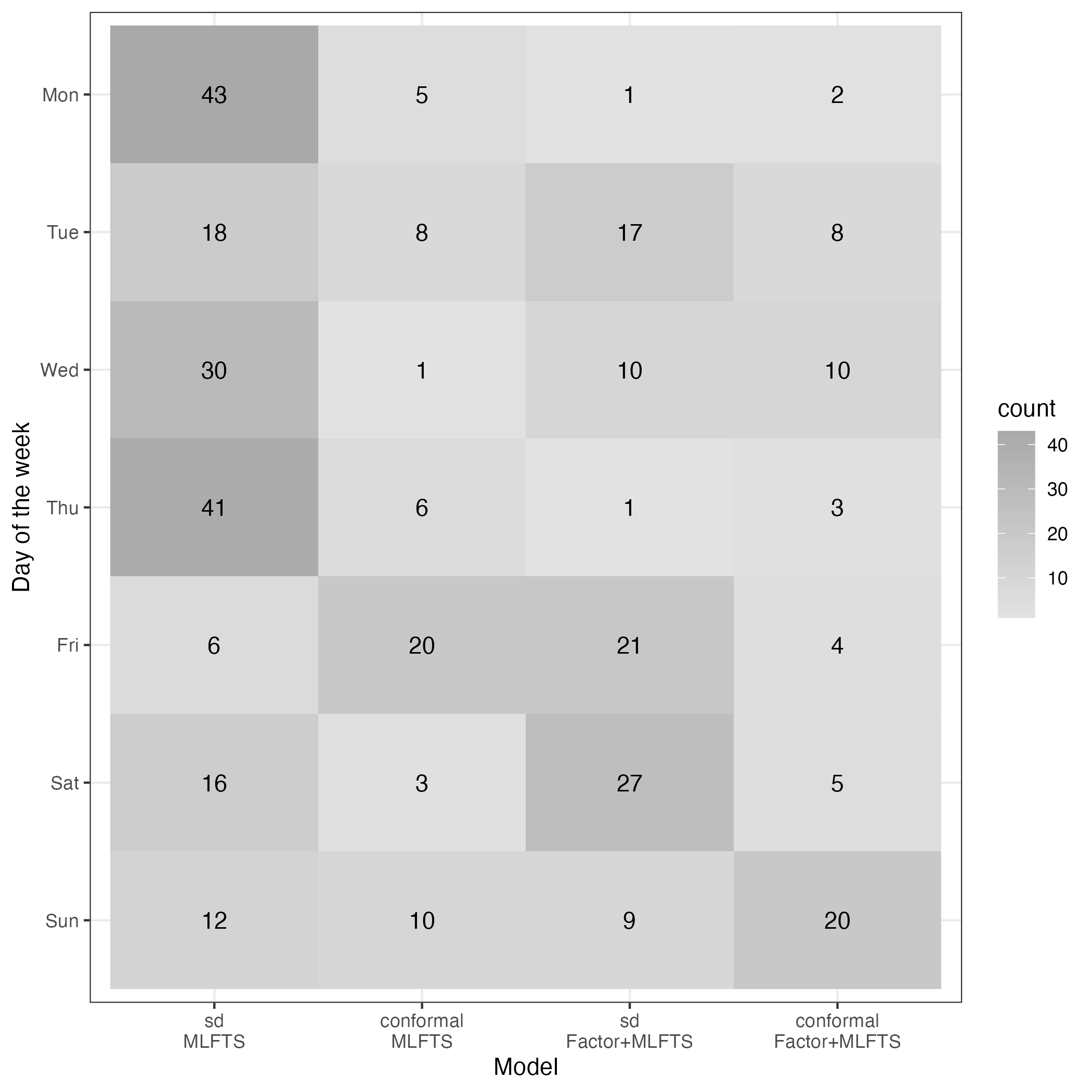}}
\caption{The CPD and mean interval scores among the functional time-series forecasting methods, comparing the sd and conformal prediction approaches.}\label{fig:7}
\end{figure}

In Figure~\ref{fig:7}, we show the number of times one method performs the best out of 51 different particle sizes for each day of the week. At the 80\% nominal coverage probability, the Factor + MLFTS method, in combination with the conformal prediction, generally provides the smallest CPD. In contrast, the MLFTS method with the conformal prediction generally provides the smallest interval scores. At the 95\% nominal coverage probability, the Factor + MLFTS method, in combination with the sd approach, generally provides the smallest CPD. In contrast, the MLFTS method with the sd approach typically provides the smallest interval scores. Between the sd and conformal prediction approaches, there is no clear recommendation.

\begin{figure}[!htb]
\centering
\subfloat[CPD]
{\includegraphics[width=8.62cm]{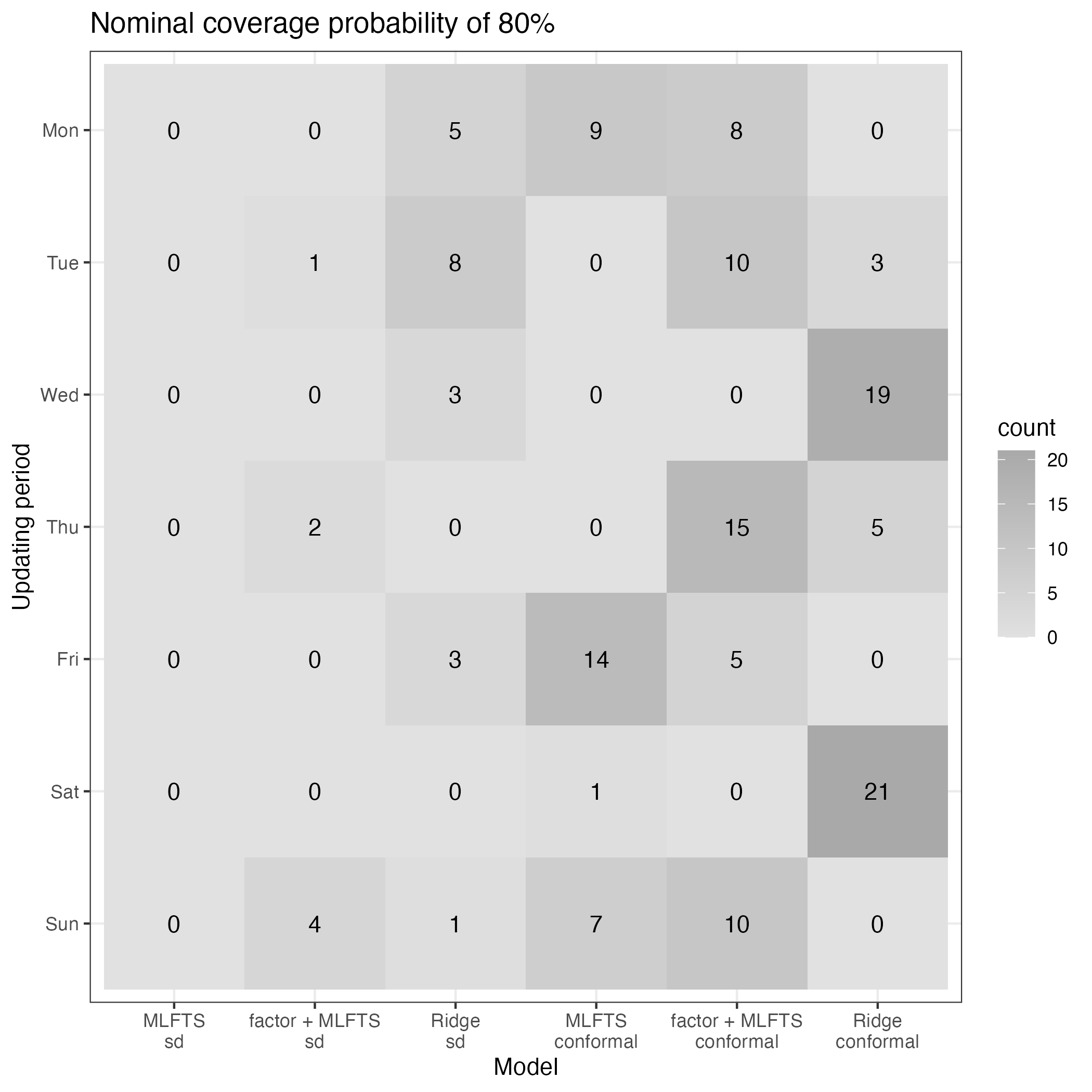}}
\quad
\subfloat[CPD]
{\includegraphics[width=8.62cm]{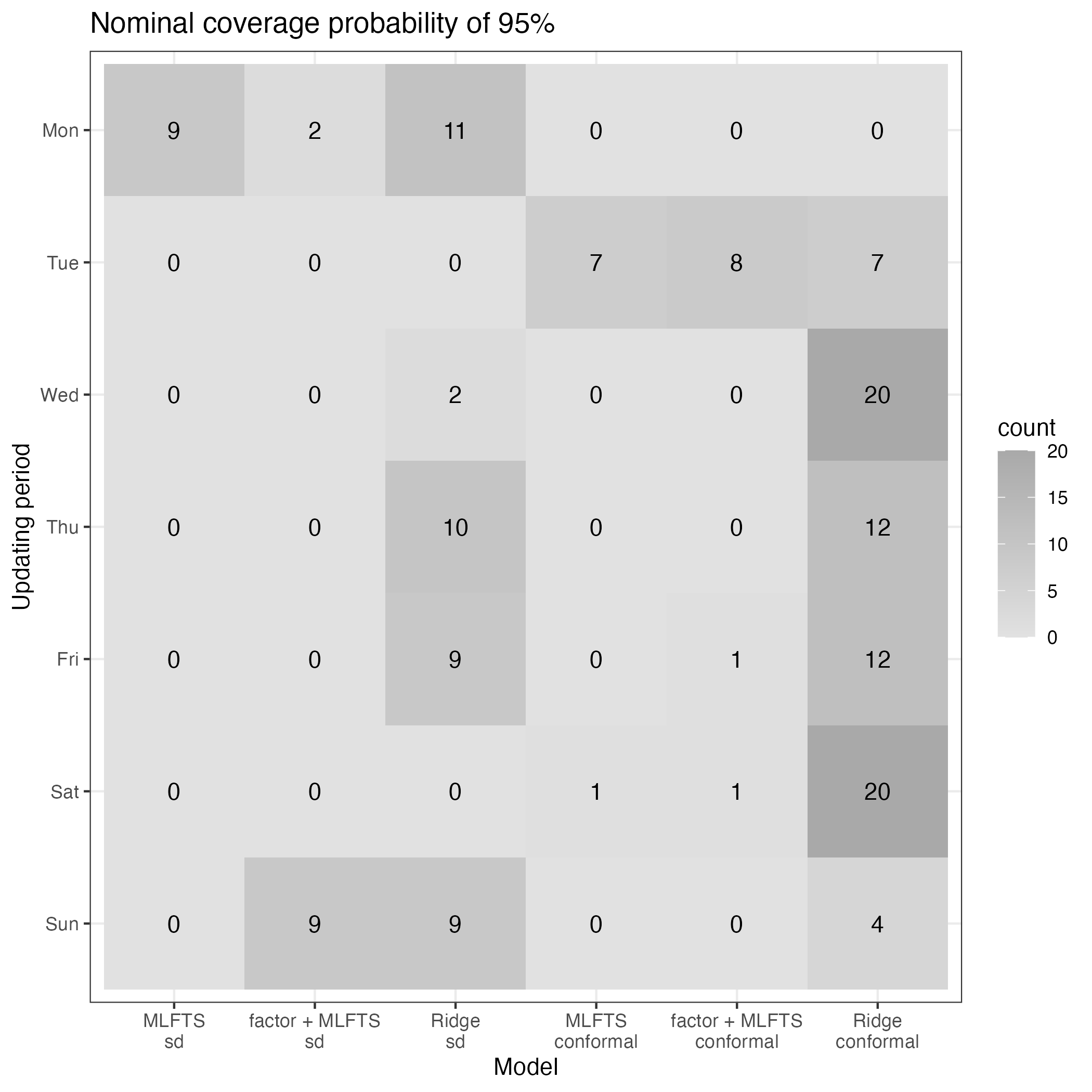}}
\\
\subfloat[Mean interval score]
{\includegraphics[width=8.62cm]{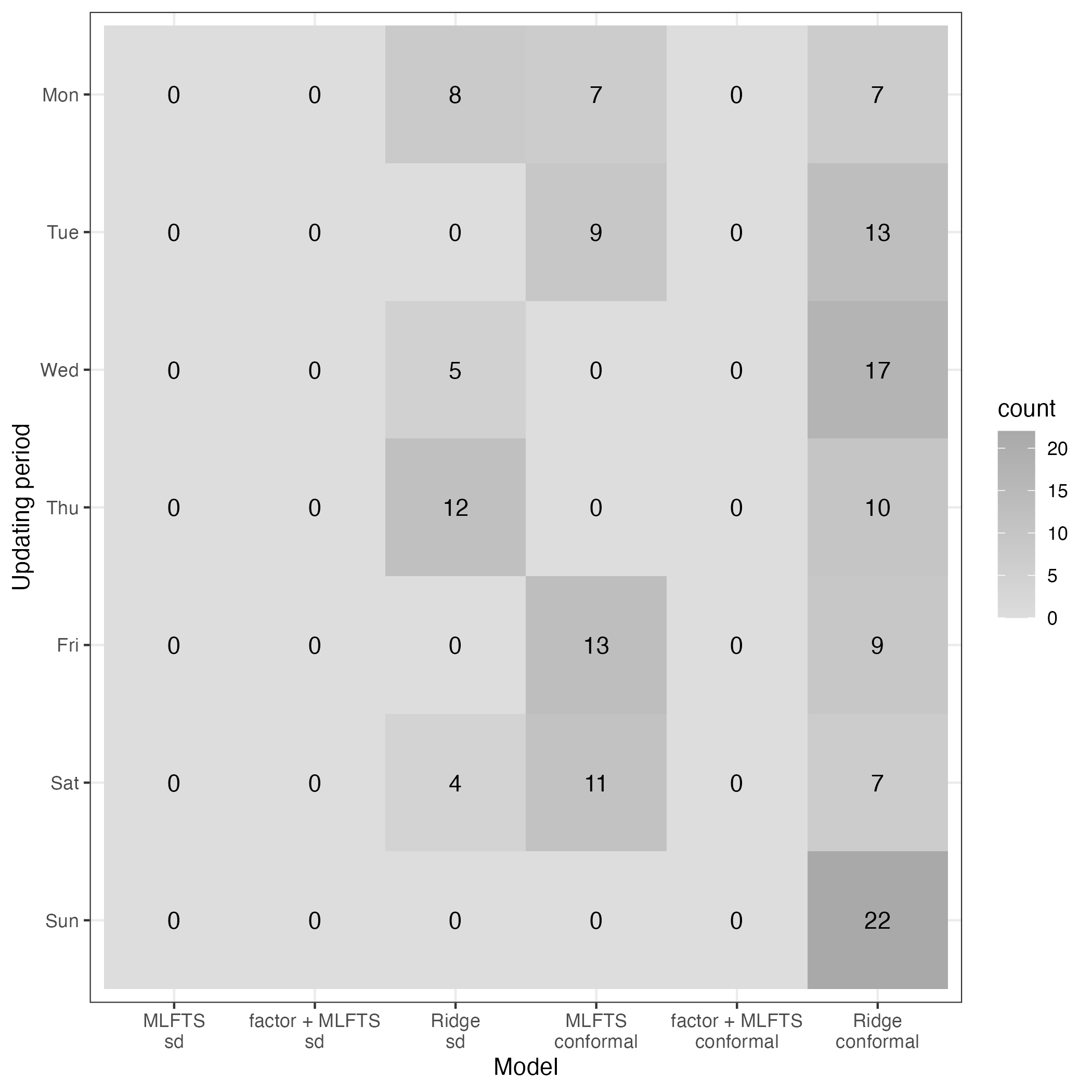}}
\quad
\subfloat[Mean interval score]
{\includegraphics[width=8.62cm]{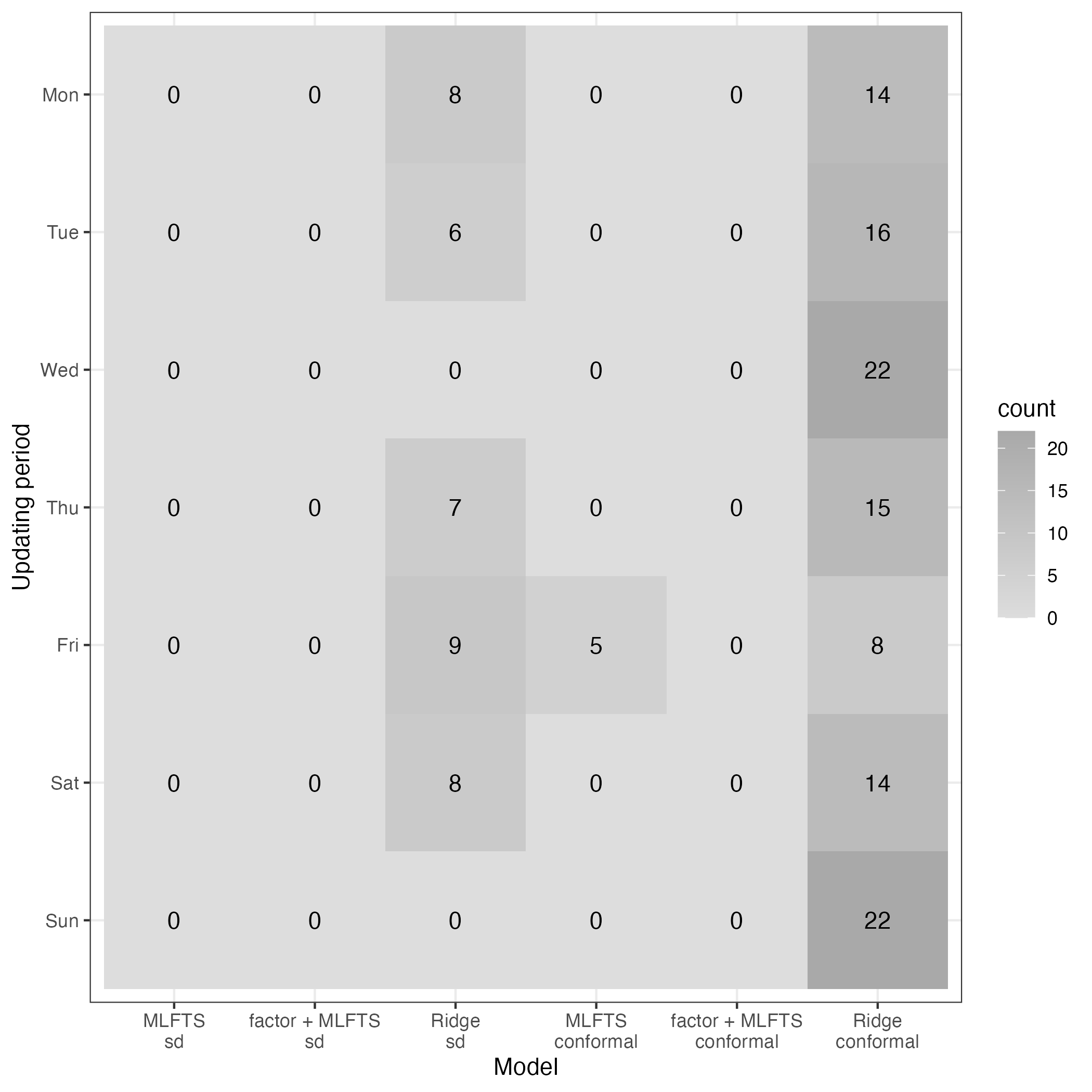}}
\caption{For 22 different updating periods, we display the frequency of the most accurate method with the smallest CPD and mean interval score for each day of the week.}\label{fig:8}
\end{figure}

When we have partially observed data in the most recent curve, we can also update our prediction intervals for the remaining period to achieve improved accuracy. Using the estimated optimal $\lambda$ values, we implement the sd and conformal prediction interval approaches to construct the prediction intervals for the remaining period. For the former method, it requires re-estimating the $\theta^s_{\alpha}$ values based on the validation dataset. In Figure~\ref{fig:8}, the ridge estimator generally provides the smallest CPD and mean interval scores. Between the sd and conformal prediction approaches, the latter seems to be more advantageous.

\section{Conclusion}\label{sec:8}

Understanding the dynamics of particulate matter of various sizes is crucial for public health. For example, it is well known that ultrafine particles (particles with diameter smaller than $100$ nm) are more harmful to health, as they cause more severe pulmonary inflammation and remain longer in the lungs. We present functional time-series methods to produce one-day-ahead point and interval forecasts. By evaluating the accuracy of the point and interval forecast, we show the superiority of the factor model in combination with the MLFTS method. While the factor model removes the trend, the MLFTS method studies the patterns in the residuals. Such residual patterns include a common pattern shared by all sizes and a specific pattern for a given size.

When intraday data are observed sequentially, it is advantageous to consider dynamic updating to improve forecast accuracy. Among dynamic updating methods, we recommend the ridge estimator and demonstrate how the estimated regression coefficient of this estimator varies over intraday periods. Furthermore, we present the sd approach and the conformal prediction approach to construct prediction intervals. For producing one-step-ahead prediction intervals, the factor model, in combination with the multilevel functional time-series method, yields the smallest CPD, albeit at the expense of larger mean interval scores compared to the multilevel functional time-series method alone. When updating interval forecasts, the ridge estimator is recommended, along with the conformal prediction approach. For reproducibility, the computer \Rlogo \ code is available at \url{https://github.com/hanshang/PNSD}.

There are several ways in which the methodology presented here can be further extended, and we briefly mention five below: 
\begin{inparaenum}
\item[1)] We study the MLFTS method, but other multiple functional time-series forecasting methods can be utilised. 
\item[2)] The presence of outliers can affect the estimation of the covariance structure, thereby affecting the accuracy of the prediction. One could consider a robust functional principal component analysis.
\item[3)] With a set of validation data, the tuning parameters in the ridge and PLS estimators can be adaptively chosen without re-computing them.
\item[4)] In the construction of prediction intervals, we compute pointwise prediction intervals. However, other standard deviations based on functional depth may be considered.
\item[5)] With continuous hourly measurements, functional data can be constructed at daily, monthly, quarterly, biannual, and annual levels. This motivates us to consider temporal forecast reconciliation of \cite{AHK+24} and \cite{GAD+24} across different data frequencies in the construction of functional data for continuously measured data sets. 
\end{inparaenum}

%\section*{Acknowledgement}

%The authors are grateful for the comments and suggestions from two reviewers, which greatly improved the manuscript. This research was supported by the Australian Research Council Future Fellowship (grant number: FT240100338).

%\newpage
\bibliographystyle{agsm}
\bibliography{PNSD.bib}

\end{document}